\documentclass[sigconf]{acmart}

\makeatletter
\def\@secfont{\sffamily\bfseries\Large}
\makeatother

\AtBeginDocument{%
  }

\usepackage[acronyms,nonumberlist,nopostdot,nomain,nogroupskip,acronymlists={hidden}]{glossaries}
\newglossary[algh]{hidden}{acrh}{acnh}{Hidden Acronyms}
\glsdisablehyper

\usepackage{tikz}
\usepackage{pgfplots}
\usepackage{glossaries}
\usepackage{soul}
\usepackage{xspace}
\usepackage{ragged2e}
\usepackage{subcaption}
\usetikzlibrary{patterns}
\usepackage{enumitem}

\newif\ifexttikz
\exttikzfalse

\ifexttikz
  \usetikzlibrary{external}
\fi

\usepackage{pgfplots}
\pgfplotsset{compat=newest}

\usepackage{enumitem}
\setlist{nosep} 

\usepackage{microtype}
\usepackage{balance} 

\newacronym{aoa}{AoA}{Angle of Arrival}
\newacronym{dapp}{dApp}{distributed Application}
\newacronym{cuphy}{cuPHY}{CUDA Physical layer}
\newacronym{cuda}{CUDA}{Compute Unified Device Architecture}
\newacronym{cubb}{cuBB}{CUDA Baseband}
\newacronym{onnx}{ONNX}{Open Neural Network Exchange}
\newacronym{ldpc}{LDPC}{Low Density Parity-Check Code}
\newacronym{crc}{CRC}{Cyclic Redundancy Check}
\newacronym{dnn}{DNN}{Deep Neural Network}
\newacronym{ort}{ORT}{ONNX Runtime}
\newacronym{tensorrt}{TensorRT}{Tensor RealTime}
\newacronym{cnn}{CNN}{Convolutional Neural Network}
\newacronym{cudnn}{cuDNN}{CUDA Deep Neural Network library}
\newacronym{cublas}{cuBLAS}{CUDA Basic Linear Algebra Subroutines}
\newacronym{hdf5}{HDF5}{Hierarchical Data Format version 5}
\newacronym{rmr}{RMR}{RIC Message Router}
\newacronym{3gpp}{3GPP}{3rd Generation Partnership Project}
\newacronym{4g}{4G}{4th generation}
\newacronym{5g}{5G}{fifth generation}
\newacronym{6g}{6G}{Sixth generation}
\newacronym{5gc}{5GC}{5G Core}
\newacronym{adc}{ADC}{Analog to Digital Converter}
\newacronym{aerpaw}{AERPAW}{Aerial Experimentation and Research Platform for Advanced Wireless}
\newacronym{ai}{AI}{Artificial Intelligence}
\newacronym{aimd}{AIMD}{Additive Increase Multiplicative Decrease}
\newacronym{am}{AM}{Acknowledged Mode}
\newacronym{amc}{AMC}{Adaptive Modulation and Coding}
\newacronym{amf}{AMF}{Access and Mobility Management Function}
\newacronym{aops}{AOPS}{Adaptive Order Prediction Scheduling}
\newacronym{api}{API}{Application Programming Interface}
\newacronym{apn}{APN}{Access Point Name}
\newacronym{aqm}{AQM}{Active Queue Management}
\newacronym{arc-ota}{ARC-OTA}{Aerial RAN CoLab Over-the-Air}
\newacronym{ausf}{AUSF}{Authentication Server Function}
\newacronym{avc}{AVC}{Advanced Video Coding}
\newacronym{awgn}{AGWN}{Additive White Gaussian Noise}
\newacronym{balia}{BALIA}{Balanced Link Adaptation Algorithm}
\newacronym{bbu}{BBU}{Base Band Unit}
\newacronym{bdp}{BDP}{Bandwidth-Delay Product}
\newacronym{ber}{BER}{Bit Error Rate}
\newacronym{bf}{BF}{Beamforming}
\newacronym{bler}{BLER}{Block Error Rate}
\newacronym{brr}{BRR}{Bayesian Ridge Regressor}
\newacronym{bsr}{BSR}{Buffer Status Report}
\newacronym{bs}{BS}{Base Station}
\newacronym{bpsk}{BPSK}{Binary Phase-shift keying}
\newacronym{bss}{BSS}{Business Support System}
\newacronym{ca}{CA}{Carrier Aggregation}
\newacronym{caas}{CaaS}{Connectivity-as-a-Service}
\newacronym{cb}{CB}{Code Block}
\newacronym{cc}{CC}{Congestion Control}
\newacronym{ccid}{CCID}{Congestion Control ID}
\newacronym{cco}{CC}{Carrier Component}
\newacronym{cdd}{CDD}{Cyclic Delay Diversity}
\newacronym{cdf}{CDF}{Cumulative Distribution Function}
\newacronym{cdn}{CDN}{Content Distribution Network}
\newacronym{cir}{CIR}{Channel Impulse Response}
\newacronym{cn}{CN}{Core Network}
\newacronym{codel}{CoDel}{Controlled Delay Management}
\newacronym{comac}{COMAC}{Converged Multi-Access and Core}
\newacronym{cord}{CORD}{Central Office Re-architected as a Datacenter}
\newacronym{cornet}{CORNET}{COgnitive Radio NETwork}
\newacronym{cosmos}{COSMOS}{Cloud Enhanced Open Software Defined Mobile Wireless Testbed for City-Scale Deployment}
\newacronym{cots}{COTS}{Commercial Off-the-Shelf}
\newacronym{cp}{CP}{Control Plane}
\newacronym{cpu}{CPU}{Central Processing Unit}
\newacronym{cqi}{CQI}{Channel Quality Information}
\newacronym{cr}{CR}{Cognitive Radio}
\newacronym{cran}{CRAN}{Cloud \gls{ran}}
\newacronym{crs}{CRS}{Cell Reference Signal}
\newacronym{csi}{CSI}{Channel State Information}
\newacronym{csirs}{CSI-RS}{Channel State Information - Reference Signal}
\newacronym{cu}{CU}{Central Unit}
\newacronym{d2tcp}{D$^2$TCP}{Deadline-aware Data center TCP}
\newacronym{d3}{D$^3$}{Deadline-Driven Delivery}
\newacronym{dac}{DAC}{Digital to Analog Converter}
\newacronym{dag}{DAG}{Directed Acyclic Graph}
\newacronym{darpa}{DARPA}{Defense Advanced Research Projects Agency}
\newacronym{das}{DAS}{Distributed Antenna System}
\newacronym{dash}{DASH}{Dynamic Adaptive Streaming over HTTP}
\newacronym{dc}{DC}{Dual Connectivity}
\newacronym{dccp}{DCCP}{Datagram Congestion Control Protocol}
\newacronym{dce}{DCE}{Direct Code Execution}
\newacronym{dci}{DCI}{Downlink Control Information}
\newacronym{dcl}{DCL}{Dear Colleague Letter}
\newacronym{dctcp}{DCTCP}{Data Center TCP}
\newacronym{dl}{DL}{Downlink}
\newacronym{dmr}{DMR}{Deadline Miss Ratio}
\newacronym{dmrs}{DMRS}{DeModulation Reference Signal}
\newacronym{drlcc}{DRL-CC}{Deep Reinforcement Learning Congestion Control}
\newacronym{drs}{DRS}{Discovery Reference Signal}
\newacronym{du}{DU}{Distributed Unit}
\newacronym{e2e}{E2E}{end-to-end}
\newacronym{e2sm}{E2SM}{E2 Service Model}
\newacronym{ecaas}{ECaaS}{Edge-Cloud-as-a-Service}
\newacronym{ecn}{ECN}{Explicit Congestion Notification}
\newacronym{edf}{EDF}{Earliest Deadline First}
\newacronym{eirp}{EIRP}{Effective Isotropic Radiated Power}
\newacronym{em}{EM}{Electro-Magnetic}
\newacronym{embb}{eMBB}{Enhanced Mobile Broadband}
\newacronym{empower}{EMPOWER}{EMpowering transatlantic PlatfOrms for advanced WirEless Research}
\newacronym{enb}{eNB}{evolved Node Base}
\newacronym{endc}{EN-DC}{E-UTRAN-\gls{nr} \gls{dc}}
\newacronym{epc}{EPC}{Evolved Packet Core}
\newacronym{eps}{EPS}{Evolved Packet System}
\newacronym{es}{ES}{Edge Server}
\newacronym{etsi}{ETSI}{European Telecommunications Standards Institute}
\newacronym[firstplural=Estimated Times of Arrival (ETAs)]{eta}{ETA}{Estimated Time of Arrival}
\newacronym{eutran}{E-UTRAN}{Evolved Universal Terrestrial Access Network}
\newacronym{faas}{FaaS}{Function-as-a-Service}
\newacronym{fapi}{FAPI}{Functional Application Platform Interface}
\newacronym{fcc}{FCC}{Federal Communications Commission}
\newacronym{fdd}{FDD}{Frequency Division Duplexing}
\newacronym{fdm}{FDM}{Frequency Division Multiplexing}
\newacronym{fdma}{FDMA}{Frequency Division Multiple Access}
\newacronym{fed4fire}{FED4FIRE+}{Federation 4 Future Internet Research and Experimentation Plus}
\newacronym{fir}{FIR}{Finite Impulse Response}
\newacronym{fit}{FIT}{Future \acrlong{iot}}
\newacronym{fpga}{FPGA}{Field Programmable Gate Array}
\newacronym{fr2}{FR2}{Frequency Range 2}
\newacronym{fs}{FS}{Fast Switching}
\newacronym{fscc}{FSCC}{Flow Sharing Congestion Control}
\newacronym{ftp}{FTP}{File Transfer Protocol}
\newacronym{fw}{FW}{Flow Window}
\newacronym{ga128}{Ga}{Golay Sequence type A}
\newacronym{ge}{GE}{Gaussian Elimination}
\newacronym{glfsr}{GLFSR}{Galois Linear Feedback Shift Register}
\newacronym{gnb}{gNB}{Next Generation Node Base}
\newacronym{gold}{Gold}{Gold}
\newacronym{gop}{GOP}{Group of Pictures}
\newacronym{gpr}{GPR}{Gaussian Process Regressor}
\newacronym{gpu}{GPU}{Graphics Processing Unit}
\newacronym{gtp}{GTP}{GPRS Tunneling Protocol}
\newacronym{gtpc}{GTP-C}{GPRS Tunnelling Protocol Control Plane}
\newacronym{gtpu}{GTP-U}{GPRS Tunnelling Protocol User Plane}
\newacronym{gtpv2c}{GTPv2-C}{\gls{gtp} v2 - Control}
\newacronym{gw}{GW}{Gateway}
\newacronym{harq}{HARQ}{Hybrid Automatic Repeat Request}
\newacronym{hetnet}{HetNet}{Heterogeneous Network}
\newacronym{hh}{HH}{Hard Handover}
\newacronym{hol}{HOL}{Head-of-Line}
\newacronym{hqf}{HQF}{Highest-quality-first}
\newacronym{hss}{HSS}{Home Subscription Server}
\newacronym{http}{HTTP}{HyperText Transfer Protocol}
\newacronym{ia}{IA}{Initial Access}
\newacronym{iab}{IAB}{Integrated Access and Backhaul}
\newacronym{ic}{IC}{Incident Command}
\newacronym{ietf}{IETF}{Internet Engineering Task Force}
\newacronym{ifw}{IFW}{Interference Free Window}
\newacronym{imsi}{IMSI}{International Mobile Subscriber Identity}
\newacronym{imt}{IMT}{International Mobile Telecommunication}
\newacronym{iot}{IoT}{Internet of Things}
\newacronym{ip}{IP}{Internet Protocol}
\newacronym{iq}{I/Q}{In-phase and Quadrature}
\newacronym{itu}{ITU}{International Telecommunication Union}
\newacronym{iw}{IW}{Interference Whitening}
\newacronym{kpi}{KPI}{Key Performance Indicator}
\newacronym{kpm}{KPM}{Key Performance Measurement}
\newacronym{kvm}{KVM}{Kernel-based Virtual Machine}
\newacronym{leo}{LEO}{Low Earth Orbit}
\newacronym{los}{LOS}{Line-of-Sight}
\newacronym{ls}{LS}{Loosely Synchronised}
\newacronym{lsm}{LSM}{Link-to-System Mapping}
\newacronym{lstm}{LSTM}{Long Short Term Memory}
\newacronym{lte}{LTE}{Long Term Evolution}
\newacronym{lxc}{LXC}{Linux Container}
\newacronym{m2m}{M2M}{Machine to Machine}
\newacronym{mac}{MAC}{Medium Access Control}
\newacronym{manet}{MANET}{Mobile Ad Hoc Network}
\newacronym{mano}{MANO}{Management and Orchestration}
\newacronym{mc}{MC}{Multi-Connectivity}
\newacronym{mcc}{MCC}{Mobile Cloud Computing}
\newacronym{mchem}{MCHEM}{Massive Channel Emulator}
\newacronym{mcs}{MCS}{Modulation and Coding Scheme}
\newacronym{mec}{MEC}{Multi-access Edge Computing}
\newacronym{mec2}{MEC}{Mobile Edge Cloud}
\newacronym{mfc}{MFC}{Mobile Fog Computing}
\newacronym{mi}{MI}{Mutual Information}
\newacronym{mib}{MIB}{Master Information Block}
\newacronym{miesm}{MIESM}{Mutual Information Based Effective SINR}
\newacronym{mimo}{MIMO}{Multiple Input, Multiple Output}
\newacronym{mgen}{MGEN}{Multi-Generator}
\newacronym{ml}{ML}{Machine Learning}
\newacronym{mlr}{MLR}{Maximum-local-rate}
\newacronym[plural=\gls{mme}s,firstplural=Mobility Management Entities (MMEs)]{mme}{MME}{Mobility Management Entity}
\newacronym{mmtc}{mMTC}{Massive Machine-Type Communications}
\newacronym{mmwave}{mmWave}{millimeter wave}
\newacronym{mpdccp}{MP-DCCP}{Multipath Datagram Congestion Control Protocol}
\newacronym{mptcp}{MPTCP}{Multipath TCP}
\newacronym{mr}{MR}{Maximum Rate}
\newacronym{mrdc}{MR-DC}{Multi \gls{rat} \gls{dc}}
\newacronym{mse}{MSE}{Mean Square Error}
\newacronym{mss}{MSS}{Maximum Segment Size}
\newacronym{mt}{MT}{Mobile Termination}
\newacronym{mtd}{MTD}{Machine-Type Device}
\newacronym{mtu}{MTU}{Maximum Transmission Unit}
\newacronym{mumimo}{MU-MIMO}{Multi-user \gls{mimo}}
\newacronym{mvno}{MVNO}{Mobile Virtual Network Operator}
\newacronym{nalu}{NALU}{Network Abstraction Layer Unit}
\newacronym{nas}{NAS}{Network Attached Storage}
\newacronym{nbiot}{NB-IoT}{Narrow Band IoT}
\newacronym{nfv}{NFV}{Network Function Virtualization}
\newacronym{nfvi}{NFVI}{Network Function Virtualization Infrastructure}
\newacronym{nic}{NIC}{Network Interface Card}
\newacronym{nlos}{NLOS}{Non-Line-of-Sight}
\newacronym{now}{NOW}{Non Overlapping Window}
\newacronym{nrdz}{NRDZ}{National Radio Dynamic Zone}
\newacronym{nsf}{NSF}{National Science Foundation}
\newacronym{nsm}{NSM}{Network Service Mesh}
\newacronym[type=hidden]{nr}{NR}{New Radio}
\newacronym{nrf}{NRF}{Network Repository Function}
\newacronym{nsa}{NSA}{Non Stand Alone}
\newacronym{nse}{NSE}{Network Slicing Engine}
\newacronym{nssf}{NSSF}{Network Slice Selection Function}
\newacronym{ntp}{NTP}{Network Time Protocol}
\newacronym{o2i}{O2I}{Outdoor to Indoor}
\newacronym{oai}{OAI}{OpenAirInterface}
\newacronym{oaic}{OAIC}{Open AI Cellular}
\newacronym{oaicn}{OAI-CN}{\gls{oai} \acrlong{cn}}
\newacronym{oairan}{OAI-RAN}{\acrlong{oai} \acrlong{ran}}
\newacronym{oam}{OAM}{Operations, Administration and Maintenance}
\newacronym[plural=\gls{obu}s,firstplural=Onboard Units (OBUs)]{obu}{OBU}{Onboard Unit}
\newacronym{ofdm}{OFDM}{Orthogonal Frequency Division Multiplexing}
\newacronym{olia}{OLIA}{Opportunistic Linked Increase Algorithm}
\newacronym{omec}{OMEC}{Open Mobile Evolved Core}
\newacronym{onap}{ONAP}{Open Network Automation Platform}
\newacronym{onf}{ONF}{Open Networking Foundation}
\newacronym{onos}{ONOS}{Open Networking Operating System}
\newacronym{oom}{OOM}{\gls{onap} Operations Manager}
\newacronym{opnfv}{OPNFV}{Open Platform for \gls{nfv}}
\newacronym{orbit}{ORBIT}{Open-Access Research Testbed for Next-Generation Wireless Networks}
\newacronym{os}{OS}{Operating System}
\newacronym{osc}{OSC}{O-RAN Software Community}
\newacronym{osm}{OSM}{Open Street Map}
\newacronym{oss}{OSS}{Operations Support System}
\newacronym{pa}{PA}{Position-aware}
\newacronym{pase}{PASE}{Prioritization, Arbitration, and Self-adjusting Endpoints}
\newacronym{pawr}{PAWR}{Platforms for Advanced Wireless Research}
\newacronym{pbch}{PBCH}{Physical Broadcast Channel}
\newacronym{pci}{PCI}{Peripheral Component Interconnect}
\newacronym{pcef}{PCEF}{Policy and Charging Enforcement Function}
\newacronym{pcfich}{PCFICH}{Physical Control Format Indicator Channel}
\newacronym{pcrf}{PCRF}{Policy and Charging Rules Function}
\newacronym{pdcch}{PDCCH}{Physical Downlink Control Channel}
\newacronym{pdcp}{PDCP}{Packet Data Convergence Protocol}
\newacronym{pdsch}{PDSCH}{Physical Downlink Shared Channel}
\newacronym{pdu}{PDU}{Packet Data Unit}
\newacronym{pdp}{PDP}{Power Delay Profile}
\newacronym{pf}{PF}{Proportional Fair}
\newacronym{pgw}{PGW}{Packet Gateway}
\newacronym{ph}{PH}{Power Headroom}
\newacronym{phich}{PHICH}{Physical Hybrid ARQ Indicator Channel}
\newacronym{phy}{PHY}{Physical}
\newacronym{pl}{PL}{Path Loss}
\newacronym{pmch}{PMCH}{Physical Multicast Channel}
\newacronym{pmi}{PMI}{Precoding Matrix Indicators}
\newacronym{powder}{POWDER}{Platform for Open Wireless Data-driven Experimental Research}
\newacronym{ppo}{PPO}{Proximal Policy Optimization}
\newacronym{ppp}{PPP}{Poisson Point Process}
\newacronym{prach}{PRACH}{Physical Random Access Channel}
\newacronym{prb}{PRB}{Physical Resource Block}
\newacronym{psnr}{PSNR}{Peak Signal to Noise Ratio}
\newacronym{pss}{PSS}{Primary Synchronization Signal}
\newacronym{pucch}{PUCCH}{Physical Uplink Control Channel}
\newacronym{pusch}{PUSCH}{Physical Uplink Shared Channel}
\newacronym{qam}{QAM}{Quadrature Amplitude Modulation}
\newacronym{qci}{QCI}{\gls{qos} Class Identifier}
\newacronym{qoe}{QoE}{Quality of Experience}
\newacronym{qos}{QoS}{Quality of Service}
\newacronym{qtgui}{QT-GUI}{QT Graphical User Interface}
\newacronym{qsfp28}{QSFP28}{Quad Small Form-factor Pluggable 28}
\newacronym{quic}{QUIC}{Quick UDP Internet Connections}
\newacronym{rach}{RACH}{Random Access Channel}
\newacronym{ran}{RAN}{Radio Access Network}
\newacronym[firstplural=Radio Access Technologies (RATs)]{rat}{RAT}{Radio Access Technology}
\newacronym{rcn}{RCN}{Research Coordination Network}
\newacronym{rdma}{RDMA}{Remote Direct Memory Access}
\newacronym{rec}{REC}{Radio Edge Cloud}
\newacronym{red}{RED}{Random Early Detection}
\newacronym{renew}{RENEW}{Reconfigurable Eco-system for Next-generation End-to-end Wireless}
\newacronym{rf}{RF}{Radio Frequency}
\newacronym{rfc}{RFC}{Request for Comments}
\newacronym{rfr}{RFR}{Random Forest Regressor}
\newacronym{ric}{RIC}{RAN Intelligent Controller}
\newacronym{near-rt-ric}{near-RT-RIC}{near-RT-\gls{ric}}
\newacronym{non-rt}{non-RT}{non-Real-Time}
\newacronym{near-rt}{near-RT}{near-Real-Time}
\newacronym{rlc}{RLC}{Radio Link Control}
\newacronym{rlf}{RLF}{Radio Link Failure}
\newacronym{rlnc}{RLNC}{Random Linear Network Coding}
\newacronym{rmse}{RMSE}{Root Mean Squared Error}
\newacronym{rnis}{RNIS}{Radio Network Information Service}
\newacronym{rr}{RR}{Round Robin}
\newacronym{rrc}{RRC}{Radio Resource Control}
\newacronym{rrm}{RRM}{Radio Resource Management}
\newacronym{rru}{RRU}{Remote Radio Unit}
\newacronym{rs}{RS}{Remote Server}
\newacronym{rsrp}{RSRP}{Reference Signal Received Power}
\newacronym{rsrq}{RSRQ}{Reference Signal Received Quality}
\newacronym{rss}{RSS}{Received Signal Strength}
\newacronym{rssi}{RSSI}{Received Signal Strength Indicator}
\newacronym{rsu}{RSU}{Road-Side Unit}
\newacronym{rtt}{RTT}{Round Trip Time}
\newacronym{ru}{RU}{Radio Unit}
\newacronym{rw}{RW}{Receive Window}
\newacronym{rx}{RX}{Receiver}
\newacronym{s1ap}{S1AP}{S1 Application Protocol}
\newacronym{sa}{SA}{standalone}
\newacronym{sack}{SACK}{Selective Acknowledgment}
\newacronym{sap}{SAP}{Service Access Point}
\newacronym{sas}{SAS}{Spectrum Access System}
\newacronym{sc2}{SC2}{Spectrum Collaboration Challenge}
\newacronym{scef}{SCEF}{Service Capability Exposure Function}
\newacronym{sch}{SCH}{Secondary Cell Handover}
\newacronym{scoot}{SCOOT}{Split Cycle Offset Optimization Technique}
\newacronym{sfp+}{SFP+}{Small Form-factor Pluggable Plus}
\newacronym{sctp}{SCTP}{Stream Control Transmission Protocol}
\newacronym{sdap}{SDAP}{Service Data Adaptation Protocol}
\newacronym{sd}{SD}{Standard Deviation}
\newacronym{sdk}{SDK}{Software Development Kit}
\newacronym{sdm}{SDM}{Space Division Multiplexing}
\newacronym{sdma}{SDMA}{Spatial Division Multiple Access}
\newacronym{sdn}{SDN}{Software-defined Networking}
\newacronym{sdr}{SDR}{Software-defined Radio}
\newacronym{seba}{SEBA}{SDN-Enabled Broadband Access}
\newacronym{sgsn}{SGSN}{Serving GPRS Support Node}
\newacronym{sgw}{SGW}{Service Gateway}
\newacronym{si}{SI}{Study Item}
\newacronym{sib}{SIB}{Secondary Information Block}
\newacronym{sinr}{SINR}{Signal to Interference plus Noise Ratio}
\newacronym{sip}{SIP}{Session Initiation Protocol}
\newacronym{siso}{SISO}{Single Input, Single Output}
\newacronym{sla}{SLA}{Service Level Agreement}
\newacronym{sm}{SM}{Saturation Mode}
\newacronym{smf}{SMF}{Session Management Function}
\newacronym{smo}{SMO}{Service Management and Orchestration}
\newacronym{sms}{SMS}{Short Message Service}
\newacronym{smsgmsc}{SMS-GMSC}{\gls{sms}-Gateway}
\newacronym{snr}{SNR}{Signal-to-Noise-Ratio}
\newacronym{son}{SON}{Self-Organizing Network}
\newacronym{sptcp}{SPTCP}{Single Path TCP}
\newacronym{srb}{SRB}{Service Radio Bearer}
\newacronym{srn}{SRN}{Standard Radio Node}
\newacronym{srs}{SRS}{Sounding Reference Signal}
\newacronym{ss}{SS}{Synchronization Signal}
\newacronym{sss}{SSS}{Secondary Synchronization Signal}
\newacronym{st}{ST}{Spanning Tree}
\newacronym{svc}{SVC}{Scalable Video Coding}
\newacronym{synce}{SyncE}{Synchronous Ethernet}
\newacronym{tb}{TB}{Transport Block}
\newacronym{tcp}{TCP}{Transmission Control Protocol}
\newacronym{tdd}{TDD}{Time Division Duplexing}
\newacronym{tdm}{TDM}{Time Division Multiplexing}
\newacronym{tdma}{TDMA}{Time Division Multiple Access}
\newacronym{tfl}{TfL}{Transport for London}
\newacronym{tfrc}{TFRC}{TCP-Friendly Rate Control}
\newacronym{tft}{TFT}{Traffic Flow Template}
\newacronym{tgen}{TGEN}{Traffic Generator}
\newacronym{tip}{TIP}{Telecom Infra Project}
\newacronym{tm}{TM}{Transparent Mode}
\newacronym{to}{TO}{Telco Operator}
\newacronym{toa}{ToA}{Time of Arrival}
\newacronym{tl}{TL}{Transfer Learning}
\newacronym{tr}{TR}{Technical Report}
\newacronym{trp}{TRP}{Transmitter Receiver Pair}
\newacronym{ts}{TS}{Technical Specification}
\newacronym{tti}{TTI}{Transmission Time Interval}
\newacronym{ttt}{TTT}{Time-to-Trigger}
\newacronym{tx}{TX}{Transmitter}
\newacronym{uas}{UAS}{Unmanned Aerial System}
\newacronym{uav}{UAV}{Unmanned Aerial Vehicle}
\newacronym{udm}{UDM}{Unified Data Management}
\newacronym{udp}{UDP}{User Datagram Protocol}
\newacronym{udr}{UDR}{Unified Data Repository}
\newacronym{ue}{UE}{User Equipment}
\newacronym{uhd}{UHD}{\gls{usrp} Hardware Driver}
\newacronym{ul}{UL}{uplink}
\newacronym{um}{UM}{Unacknowledged Mode}
\newacronym{uml}{UML}{Unified Modeling Language}
\newacronym{upa}{UPA}{Uniform Planar Array}
\newacronym{upf}{UPF}{User Plane Function}
\newacronym{urllc}{URLLC}{Ultra Reliable and Low Latency Communication}
\newacronym{usa}{U.S.}{United States}
\newacronym{usim}{USIM}{Universal Subscriber Identity Module}
\newacronym{usrp}{USRP}{Universal Software Radio Peripheral}
\newacronym{utc}{UTC}{Urban Traffic Control}
\newacronym{vim}{VIM}{Virtualization Infrastructure Manager}
\newacronym{vlan}{VLAN}{Virtual Local Area Network}
\newacronym{vm}{VM}{Virtual Machine}
\newacronym{vnf}{VNF}{Virtual Network Function}
\newacronym{volte}{VoLTE}{Voice over \gls{lte}}
\newacronym{voltha}{VOLTHA}{Virtual OLT HArdware Abstraction}
\newacronym{vr}{VR}{Virtual Reality}
\newacronym{vran}{vRAN}{Virtualized \gls{ran}}
\newacronym{vss}{VSS}{Video Streaming Server}
\newacronym{wbf}{WBF}{Wired Bias Function}
\newacronym{wf}{WF}{Wired-first}
\newacronym{wi}{WI}{Wireless InSite}
\newacronym{wlan}{WLAN}{Wireless Local Area Network}
\newacronym{pnf}{PNF}{Physical Network Function}
\newacronym{drl}{DRL}{Deep Reinforcement Learning}
\newacronym{mtc}{MTC}{Machine-type Communications}
\newacronym{v2x}{V2X}{Vehicle-to-everything}
\newacronym{cast}{\textit{CaST}}{Channel emulation generator and Sounder Toolchain}
\newacronym{abr}{ABR}{Adaptive Bitrate Streaming}
\newacronym{arc}{ARC}{Aerial RAN CoLab}
\newacronym{dsp}{DSP}{Digital Signal Processing}
\newacronym{ota}{OTA}{Over-the-Air}
\newacronym{bom}{BoM}{Bill of Materials}
\newacronym{frand}{FRAND}{Fair, Reasonable, And Non-Discriminatory}
\newacronym{nvipc}{NVIPC}{NVIDIA Inter-Process Communication}
\newacronym{hdr}{HDR}{High Dynamic Range}
\newacronym{ipc}{IPC}{Inter-Process Communication}
\newacronym{uci}{UCI}{Uplink Control Indication}
\newacronym{cbrs}{CBRS}{Citizen Broadband Radio Service}
\newacronym{ptp}{PTP}{Precision Timing Protocol}
\newacronym{scf}{SCF}{Small Cell Forum}
\newacronym{re}{RE}{Resource Element}
\newacronym{fp16}{FP16}{Float16}
\newacronym{fp32}{FP32}{Float32}
\newacronym{int32}{INT32}{32-bit Integer}
\newacronym{tdl}{TDL}{Tapped Delay Line}
\newacronym{fh}{FH}{Front-haul}
\newacronym{ta}{TA}{Timing Advance}
\newacronym{cfo}{CFO}{Carrier Frequency Offset}
\newacronym{sir}{SIR}{Signal to Interference Ratio}
\newacronym{mmse}{MMSE}{Minimum Mean Square Error}
\newacronym{irc}{IRC}{Interference Rejection Combining}
\newacronym{tf}{TF}{Tensorflow}
\acmDOI{XXXXXXX.XXXXXXX}

\acmConference[MobiHoc '25]{MobiHoc '25}{Oct 27--30,
  2025}{Houston, Texas, US}
\acmISBN{978-1-4503-XXXX-X/18/06}

\newcommand{\dapp}{InterfO-RAN\xspace}

\usepackage{tikzpagenodes,etoolbox}
\usetikzlibrary{calc}
\usepackage[contents={}]{background}
\AddEverypageHook{%
\ifnumequal{\thepage}{1}{%
    \tikz[remember picture,overlay]{%
        \node[draw,
        minimum width=1.03\textwidth,
        text width=1.02\textwidth,
        font=\footnotesize
        ]
        at ($(current page header area) - (0,-15pt)$)
        {%
        This paper has been accepted for publication on the Proc. of ACM MobiHoc 2025. This is the author's accepted version of the article. The final version published by ACM is N. Neasamoni Santhi, D. Villa, M. Polese, T. Melodia, \textit{``InterfO-RAN: Real-Time In-band Cellular Uplink Interference Detection with GPU-Accelerated dApps,``} Proceeding of ACM International Symposium on Theory, Algorithmic Foundations, and Protocol Design for Mobile Networks and Mobile Computing (MobiHoc), Houston, TX, USA, October 2025.
        };
    }%
}{}
}

\pgfplotsset{compat=1.18} 
\begin{document}
\title{\dapp: Real-Time In-band Cellular Uplink Interference Detection with GPU-Accelerated dApps}

\author{Neagin Neasamoni Santhi, Davide Villa, Michele Polese, Tommaso Melodia}
\affiliation{%
  \country{Institute for the Wireless Internet of Things, Northeastern University, Boston, MA, U.S.A \\ \{neasamonisanthi.n, villa.d, m.polese, t.melodia\}@northeastern.edu}
}

\begin{abstract}

Ultra-dense \gls{5g} and beyond networks leverage spectrum sharing and frequency reuse to enhance throughput, but face unpredictable in-band \gls{ul} interference challenges that significantly degrade \gls{sinr} at affected \glspl{gnb}. This is particularly problematic at cell edges, where overlapping regions force \glspl{ue} to increase transmit power, and in directional millimeter wave systems, where beamforming sidelobes can create unexpected interference. The resulting signal degradation disrupts protocol operations, including scheduling and resource allocation, by distorting quality indicators like \gls{rsrp} and \gls{rssi}, and can compromise critical functions such as channel state reporting and \gls{harq} acks.

To address this problem, this article introduces \dapp, a real-time programmable solution that leverages a \gls{cnn} to process \gls{iq} samples in the \gls{gnb} physical layer, detecting in-band interference with accuracy exceeding 91\% in under $650$ $\mu$s. \dapp represents the first O-RAN dApp accelerated on \gls{gpu}, coexisting with the 5G NR physical layer processing of NVIDIA Aerial. Deployed in an end-to-end private 5G network with commercial \glspl{ru} and smartphones, our solution was trained and tested on more than $7$ million \gls{nr} \gls{ul} slots collected from real-world environments, demonstrating robust interference detection capabilities essential for maintaining network performance in dense deployments.

\end{abstract}


\newlength\fwidth
\newlength\fheight

\glsresetall
\glsunset{nr}

\makeatletter
\def\@ACM@copyright@check@cc{}
\makeatother

\acmYear{2025}\copyrightyear{2025}
\setcopyright{cc}
\setcctype[4.0]{by-nc-sa}
\acmConference[MobiHoc '25]{International Symposium on Theory, Algorithmic Foundations, and Protocol Design for Mobile Networks and Mobile Computing}{October 27--30, 2025}{Houston, TX, USA}
\acmBooktitle{International Symposium on Theory, Algorithmic Foundations, and Protocol Design for Mobile Networks and Mobile Computing (MobiHoc '25), October 27--30, 2025, Houston, TX, USA}
\acmDOI{10.1145/3704413.3764454}
\acmISBN{979-8-4007-1353-8/25/10}
\maketitle

\glsresetall
\glsunset{nr}
\glsunset{5g}
\glsunset{dapp}

\section{Introduction}
\label{sec:intro}

The evolution of wireless communication networks 
has been driven by the increasing demand for high-speed, low-latency, and ultra-reliable connectivity to support emerging applications such as autonomous systems, industrial automation, and immersive experiences. To meet these demands, 5G-and-beyond networks leverage massive densification to improve throughput and latency within the limited spectrum allocated to mobile communications~\cite{ultradense}. 
In ultra-dense networks, multiple neighboring small cells are configured to reuse the same frequency bands to increase frequency reuse~\cite{ultradense5g}. 
Directional millimeter wave systems further push this concept, with commercial deployments often leveraging the same $400$~MHz or $800$~MHz bands at $28$~GHz and $39$~GHz for all base stations~\cite{9796693}.
Finally, private network deployments (e.g., for enterprise scenarios) use limited portions of shared spectrum, with all deployments constrained in the same $100$~MHz (e.g., Germany, Brazil, Netherlands) or $150$~MHz (e.g., U.S., with \acrshort{cbrs})~\cite{spec_availability}.

\begin{figure}[b]
  \centering
   \includegraphics[scale=.47]{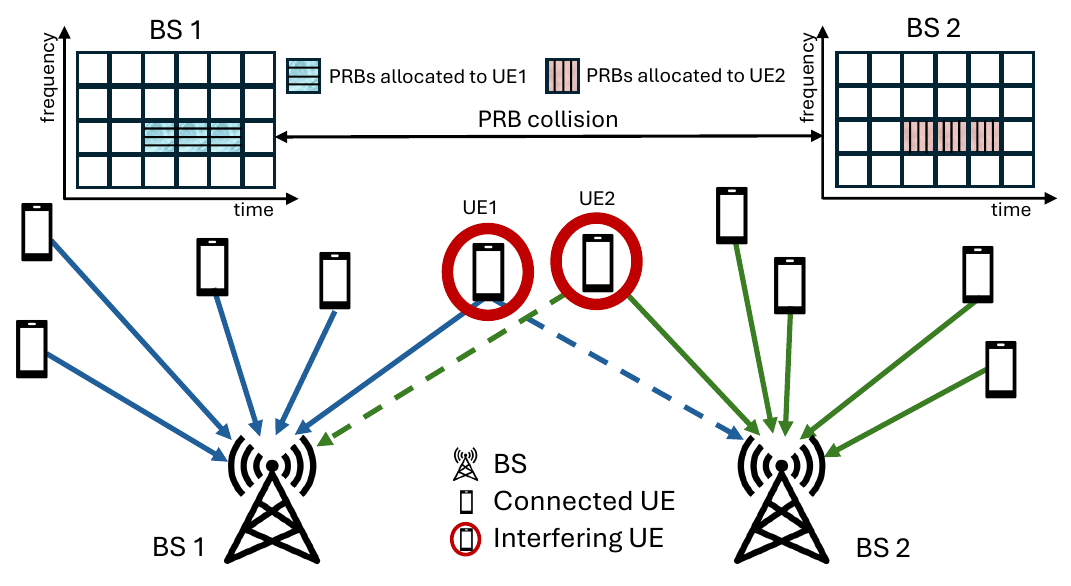}
\setlength\abovecaptionskip{-.1cm}
  \caption{In-band \justifying \acrshort{ul} interference overview (dashed lines).}
  \label{fig:interfernce_schematic}
  \Description{Inference time between standalone and cuBB for different execution providers.}
\end{figure}

\textbf{The Need for Real-Time Interference Detection.}
However, densification, sharing, and spectrum reuse also increase inter-cell interference. Although significant research has focused on coordination mechanisms to reduce interference in downlink~\cite{8938771,9405679,9411723,7536929},
in-band \gls{ul} interference remains a challenging problem, especially considering the unpredictability of user mobility and configurations and thus of the source of interference. This occurs when unwanted transmissions from \glspl{ue}, operating within the same frequency band but connected to different \glspl{gnb}, interfere with the \gls{ul} reception of a serving \gls{gnb}, as shown in Fig.~\ref{fig:interfernce_schematic}. Such an overlap can significantly degrade the \gls{ul} \gls{sinr} at the affected \gls{gnb}, potentially leading to unrecoverable packets and thus reduced network performance~\cite{7127550}.
This is particularly significant when \glspl{ue} are located at the cell edge, where overlapping regions between two neighboring cells force the devices to increase their transmit power to overcome the link budget constraints. 
In such cases, the unpredictability of potential in-band \gls{ul} interference is closely associated with \gls{ue} power control. Similarly, \glspl{ue} may transmit at excessive power levels to compensate for signal degradation when experiencing \gls{nlos} conditions with respect to their serving node. While this adjustment helps maintain reliable links, it can also lead to unintended interference in adjacent cells. 
Moreover, in scenarios involving directional \gls{ue} transmissions, as in millimeter wave links, beamforming introduces an additional source of unpredictability. It can impact adjacent cells through strong sidelobes, or, under highly dynamic conditions, even affect them with the main lobe. These conditions are often unknown apriori (e.g., at scheduling time), and can vary on timescales faster than a subframe (i.e., sub-ms). This calls for real-time interference detection to enable a prompt reaction across the protocol stack.

The impact of in-band \gls{ul} interference extends beyond immediate signal degradation, and can disrupt protocol operations like scheduling and resource allocation at the \gls{gnb}. To execute these operations, the \gls{gnb} relies on quality indicators like \gls{rsrp} and \gls{rssi}. However, interference distorts the \gls{gnb}'s assessment of \gls{ue} channel conditions, introducing errors that propagate across multiple slots and affect both \gls{ul} and \gls{dl} transmissions. This disruption becomes particularly critical when interference affects control signaling (e.g., \gls{uci}) or \gls{dmrs} used for channel estimation. 
Because of the inherently lower power of \gls{ul} signals compared to \gls{dl} transmissions, even interference at low power levels can significantly impair communication quality. This forces the \gls{gnb} to allocate additional resources through retransmissions and more robust coding schemes, thereby reducing spectral efficiency and limiting overall user capacity. In scenarios with severe interference, connection drops can occur, underscoring the importance of developing and implementing robust interference detection and mitigation methods in practical deployments.

\textbf{Contribution: \dapp.} In this paper, we address the critical challenge of \gls{ul} {\bf interference detection and mitigation} by presenting \dapp, a novel real-time programmable solution that integrates a \gls{cnn} within the \gls{gnb} physical layer to process and analyze \gls{iq} samples. Our solution achieves interference detection with accuracy exceeding $91$\% in less than $650$ $\mu$s, enabling practical implementation in production cellular networks.

Designed as a pluggable, programmable component, \dapp is the first to extend \gls{gpu} acceleration to \glspl{dapp} for real-time \gls{cnn}-based interference detection. The \dapp dApp is integrated with the NVIDIA Aerial \gls{5g} \gls{nr} \gls{phy} layer following the emerging O-RAN \gls{dapp} paradigm~\cite{lacava2025dapps}. This represents a fundamental system-level innovation, enabling pluggable \gls{ai}/\gls{ml} models for \gls{5g} and beyond and accelerated on \gls{gpu}, aligning with AI-for-RAN use cases within the AI-RAN Alliance~\cite{airan}. \dapp's design requires \gls{gpu} resource management, seamless interfacing between the \gls{dapp} inference (using \gls{ort}) and the \gls{phy}, and careful model design (quantized, minimal, yet effective \gls{cnn}). The \gls{dapp} accesses \gls{ul} data (\gls{iq} samples and channel quality \glspl{kpi}) and telemetry and provides a reusable, extensible interface for additional real-time \gls{ml}-based functionalities in the O-RAN stack. The same \gls{dapp}, indeed, can be repurposed for tasks like \gls{aoa} Estimation or 5G positioning.

\dapp is implemented as a set of pipelines to expose necessary information from Aerial \gls{cuda} kernels related to the \gls{pusch}, as well as additional \gls{dapp} \gls{cuda} kernels that implement \gls{cnn} with \gls{ort}. We selected \glspl{cnn} as they effectively recognize patterns in complex \gls{iq} planes. We deploy \dapp and the Aerial stack in an end-to-end private \gls{5g} network. The private \gls{5g} network is a multi-vendor, fully virtualized, and \gls{gpu}-accelerated production network with multiple commercial \glspl{ru} and smartphones~\cite{villa2024x5g}, where the higher layers of the \gls{5g} stack are implemented with \gls{oai}. We leverage this real-world setup to collect data with and without interference from a deployment spanning two different buildings. More than seven million \gls{nr} \gls{ul} slots, along with artificially generated data in MATLAB, are used to train and test eight configurations of the \gls{cnn}, employing \gls{tl} techniques as well. The selected solution is then evaluated \gls{ota} across various scenarios with different pairs of interfering \glspl{ru}. Our results demonstrate the robustness of \dapp in detecting interference, achieving over 91\% accuracy in less than $650~\mu$s, while imposing minimal strain on \gls{phy} operations.

The remainder of the paper is organized as follows. Section~\ref{sec:system} presents the system architecture overview. Section~\ref{sec:implementation-prelim} provides some background on the Aerial framework. Section~\ref{sec:dapp} describes the system design and implementation of the \dapp dApp, including details on the \gls{cnn} and data pipelines. Section~\ref{sec:framework} outlines the experimental setup and data collection operations, while Sec.~\ref{sec:results} discusses the results. Related work is reviewed in Sec.~\ref{sec:relatedwork}. Finally, Sec.~\ref{sec:conclusion} concludes the paper and outlines directions for future research.

\begin{figure}[t]
  \centering
    \includegraphics[scale=.45]{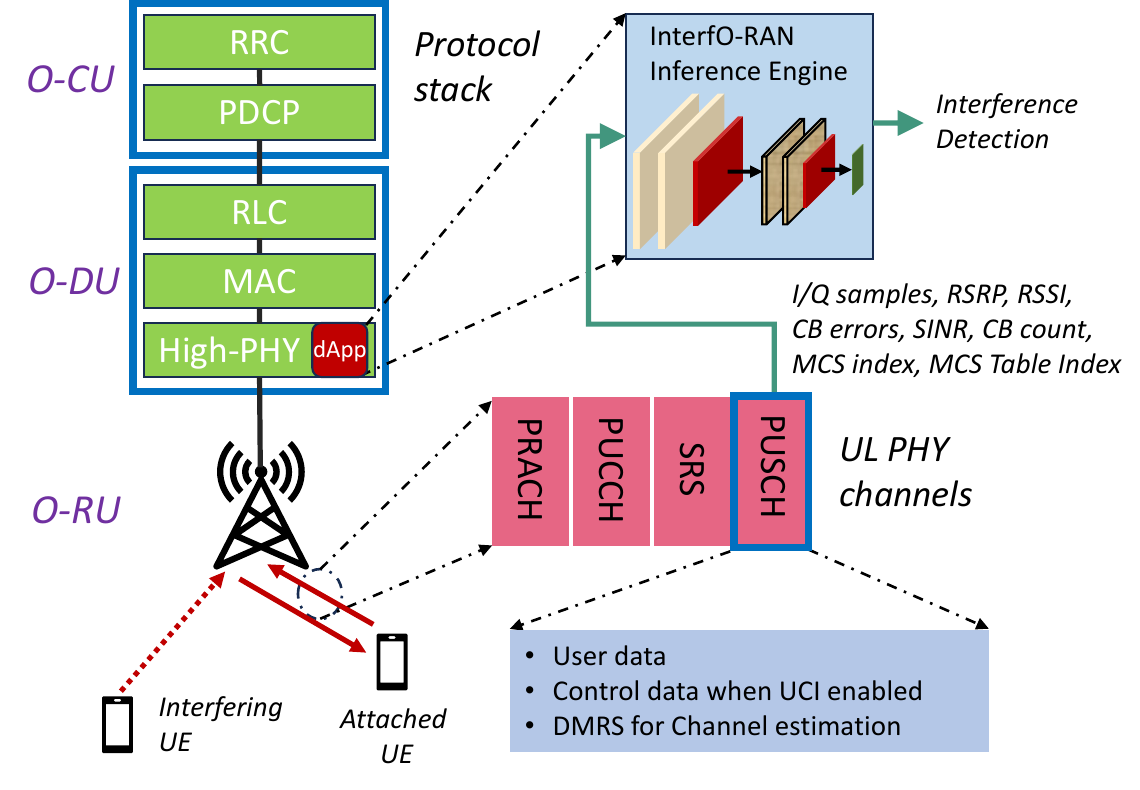}
    \setlength\belowcaptionskip{-.3cm}
    \setlength\abovecaptionskip{-.3cm}
  \caption{\justifying System Architecture.}
  \label{fig:system_archi}
  \Description{System Architecture.}
\end{figure} 

\section{System Architecture}
\label{sec:system}

\textbf{O-RAN Primer.} The system model of this paper develops on top of the O-RAN architecture. O-RAN disaggregates the traditional monolithic \gls{gnb} into modular, programmable components using open interfaces, i.e., the \gls{ru}, \gls{du}, and \gls{cu}, as shown in Fig.~\ref{fig:system_archi}.
Within this framework, the features and capabilities of the \gls{ran} can be extended with custom logic that interfaces with the disaggregated \gls{gnb}, implementing control loops for inference, control, and classifications. Among these, 
dApps are distributed applications designed to bring intelligence to \glspl{cu} and \glspl{du}, supporting real-time inference and control loops operating at sub-$10$~ms timescales~\cite{lacava2025dapps}. dApps complement xApps and rApps residing within external \glspl{ric}~\cite{polese2023understanding}.

\textbf{The \dapp Framework.}
Figure~\ref{fig:system_archi} illustrates our system architecture, where \dapp is integrated as a programmable dApp into the high-\gls{phy} of a \gls{5g} \gls{nr} protocol stack. 
%
Specifically, we design the system to harness \gls{gpu} resources available as part of the NVIDIA Aerial \gls{gpu}-accelerated platform. 
Aerial already uses \gls{gpu} to offload computationally intensive and time-sensitive high-\gls{phy} layer workloads, contributing approximately $85$\% of the overall computational complexity and processing demands in \glspl{gnb}~\cite{kundu2023hardwareaccelerationopenradio}.
This architecture enables efficient resource sharing, allowing the dApp to perform inference with minimal computational overhead while preserving real-time processing capabilities. 
As explained in Sec.~\ref{sec:dapp}, \dapp is implemented through customized functions utilizing a \gls{cnn} architecture, executed on a \gls{gpu} for real-time interference detection. The output of \dapp is a binary indicator that denotes the presence of interference. Once interference is detected, different strategies can be put in place for mitigation (e.g., coordinated resource allocation), which are left for future work.

The system processes input features comprising a combination of raw \gls{iq} samples within a slot carrying \gls{pusch} data, along with \gls{rssi}, \gls{rsrp}, the number of \gls{cb} errors, the total count of \glspl{cb}, \gls{sinr}, \gls{mcs} index and \gls{mcs} table index. The \gls{iq} samples are extracted from the \gls{pusch} physical channel, without any processing such as equalization, demodulation, or decoding. \dapp does not interfere with ongoing \gls{ul}/\gls{dl} processes, let alone \gls{pusch} channel processes, while determining whether transmissions are affected by unwanted signals within the same frequency band.

Figure~\ref{fig:system_archi} further illustrates the \gls{ul} interaction between \gls{gnb} and \gls{ue}, providing a comprehensive view of the data flow and the \gls{pusch} channel engaged by \dapp. The \gls{pusch} carries data, control information, and \gls{dmrs} for channel estimation. The figure also highlights the other uplink channels, i.e., \gls{pucch}, \gls{srs}, and \gls{prach}.

In the next sections, we provide a preliminary overview of the GPU-accelerated framework for physical layer processing, and then detail how \dapp is designed and implemented in detail.

\vspace{-5pt}
\section{GPU-Accelerated Physical Layer Processing}
\label{sec:implementation-prelim}

This section provides a detailed overview of the framework we leverage as the basis for the \dapp implementation, as well as notions on the physical layer \gls{pusch}.

\subsection{NVIDIA Aerial Physical Layer}
\label{subsec:cubb}

\dapp is embedded as a functional plugin in NVIDIA Aerial \gls{cubb}, a \gls{sdk} developed by NVIDIA that provides a \gls{5g} signal processing pipeline for \glspl{gpu}, implemented in \gls{cuda} and C++~\cite{arc-ota}. \gls{cubb} operates with slot-level granularity, with each slot lasting $500$~$\mu$s, aligned with a $30$~kHz subcarrier spacing. \dapp complements \gls{cubb} with real-time, high-performance inference.
NVIDIA Aerial uses a \gls{gpu} for inline acceleration of resource-intensive tasks such as channel estimation, \gls{ldpc} encoding/decoding, and channel equalization, among others.

From an implementation standpoint, \gls{cubb} combines several software components, which we extend to design and implement \dapp. The \gls{cuphy} controller serves as the primary coordinator, initializing \gls{gpu} resources and creating the initial context for \gls{ru} connections. During \gls{ota} \gls{ul}/\gls{dl} transmissions, the L2 adapter within the \gls{cuphy} controller translates control plane messages from the \gls{scf} \gls{fapi} interface into slot commands for \gls{phy}-layer data traffic. These commands are processed by the \gls{cuphy} controller, converted into tasks, and assigned to specific \gls{ul} and \gls{dl} channel worker threads. Mapped to appropriate \gls{cpu} cores, these threads delegate computational tasks to the \gls{gpu}. Each thread represents a physical channel implemented as a pipeline, with operations executed on \gls{cuda} kernels. \gls{cuda} kernels are functions executed in parallel on a \gls{gpu}. The pipeline is supported by \glspl{api} provided by the \gls{cuphy} library, invoked by the \gls{cuphy} driver to manage creation, configuration, and execution. 

\vspace{-10pt}

\subsection{\gls{pusch} Operations}
\label{subsec:pusch-cubb}

\dapp focuses on the \gls{ul} receive path, particularly the \gls{pusch}, to detect interference. \gls{ul} processing starts with slot configuration from the L2 layer and ends with \gls{ul} signal processing. Among \gls{ul} channels (\gls{pusch}, \gls{pucch}, \gls{prach}, \gls{srs}), \dapp analyzes the \gls{pusch}, the primary \gls{ul} channel for data transmission, to detect interference from other \glspl{ue} sharing the same radio resources. 

\gls{pusch} transmissions are dynamically scheduled by the \gls{gnb}, which allocates frequency and time resources based on real-time network conditions. The resource allocation is signaled to the \gls{ue} through \gls{dci} messages, specifying parameters such as \gls{mcs}, resource block allocation, and transmission power~\cite{dahlman2020}.
The \gls{pusch} channel in \gls{cubb} is implemented as a pipeline, for operations such as \gls{re} demapping, channel estimation, channel equalization, de-rate matching, de-layer mapping, de-scrambling, \gls{ldpc} decoding, and \gls{crc} checks for both \glspl{cb} and \glspl{tb}. Additionally, it includes processing for \gls{uci} on \gls{pusch} and transform precoding, if enabled, executed through individual \gls{cuda} kernels. Specialized kernels also manage \gls{ta} and \gls{cfo}.

\begin{figure}[htbp]
  \centering
    \includegraphics[scale=.46]{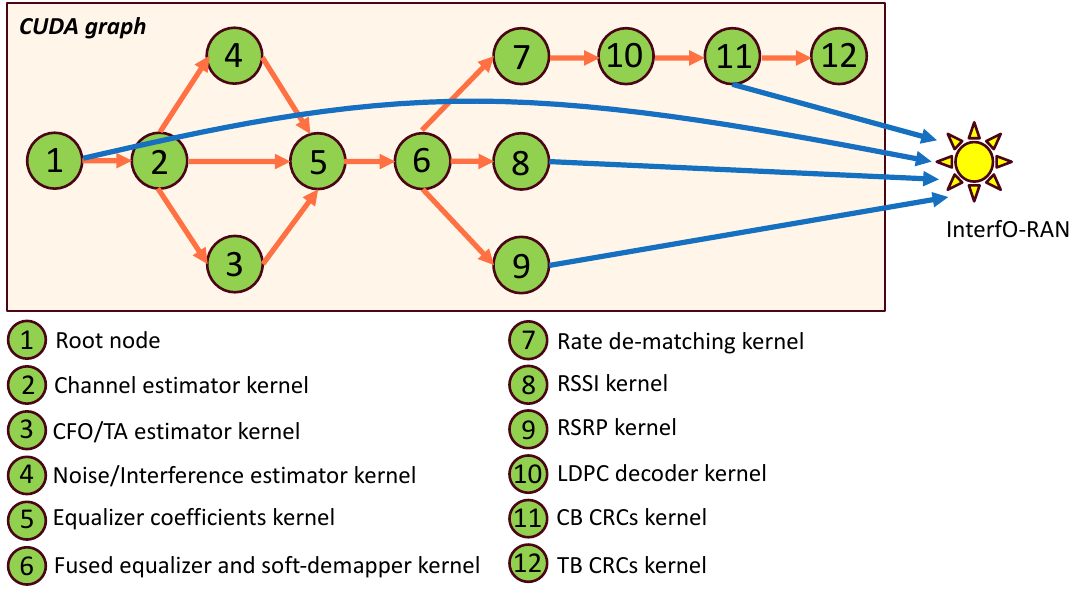}
    \captionsetup{aboveskip=.1pt, belowskip=.1pt} 
  \caption{\justifying \dapp as a post-processing component following the PUSCH CUDA graph, where each node represents a CUDA kernel.}
  \label{cuda_graph}
\end{figure}

This pipeline is further optimized and structured using an advanced feature known as \gls{cuda} graph, which represents an acyclic graph where nodes correspond to kernels, and edges define their inter-dependencies, as illustrated in Fig.~\ref{cuda_graph}. This graph-based architecture effectively delineates the intricate workflow of kernel operations, enabling the management of dependencies within the graph itself. By facilitating the concurrent launch of all kernels as a unified entity, the \gls{gpu} autonomously oversees the execution process, thereby reducing the necessity for continuous \gls{cpu} intervention. This method significantly enhances temporal efficiency by minimizing launch overhead, which is the latency associated with initiating \gls{cuda} kernels.

\section{\dapp Design and Implementation}
 \label{sec:dapp}

We design and implement \dapp on top of the NVIDIA Aerial \gls{sdk}, leveraging the spare capacity of the \gls{gpu} to execute the \gls{cnn}-based dApp.
As discussed in detail in the results analysis in Sec.~\ref{sec:results}, the typical execution of the \gls{phy} layer in \gls{cubb} utilizes at most $50$\% of the \gls{gpu} resources, leaving substantial computational capacity available for additional tasks. This excess capacity enables \dapp to take advantage of the remaining \gls{gpu} resources for interference detection.
In doing this, however, it becomes necessary to design the dApp to efficiently access the available resources and to avoid interfering with real-time processing constraints of the base station physical layer. 

\vspace{-10pt}
\subsection{dApp Design and PHY Integration}
\label{sec:dappdesign}

\textbf{Design and Implementation Challenges.} To design and integrate \dapp within the 5G NR GPU-accelerated PHY, we addressed the following challenges. First, the dApp needs to access information available across different processing steps of the \gls{pusch} pipeline (i.e., \gls{iq} samples, \gls{rsrp}, \gls{rssi}, the sum of \gls{cb} \gls{crc} errors, count of \glspl{cb} derived from \gls{cb} \glspl{crc}, \gls{sinr}, \gls{mcs} Index and \gls{mcs} table index). This can amount to 183,484 bytes for each \gls{ul} transport block, thus requiring an efficient mechanism to expose such information. Second, the \gls{cnn} needs to return results in less than a millisecond (i.e., a 5G NR subframe) to make sure that the output is relevant to take further decisions across the stack. Therefore, the GPU-based implementation needs to be efficient in running inference with the provided input. In addition, there may be a need to support different models or configurations for the \gls{ai} processing, thus managing the lifecycle of the overall model. 
For these reasons, we select \gls{ort} as the provider, making it possible to efficiently deploy trained models on the GPU-based dApp. At the same time, \gls{ort} requires a significant loading and configuration time the first time it is executed. In general, as part of our design, it is important to avoid disrupting operations of the rest of the physical layer, thus guaranteeing that the timing of any operation within the dApp does not affect physical layer processing. 
Finally, for development purposes, the \dapp implementation needs to enable automated data collection and labeling, to streamline the gathering of samples for the training of \gls{ai} models.

\begin{figure*}[htbp]
  \centering
  \includegraphics[scale=.535]{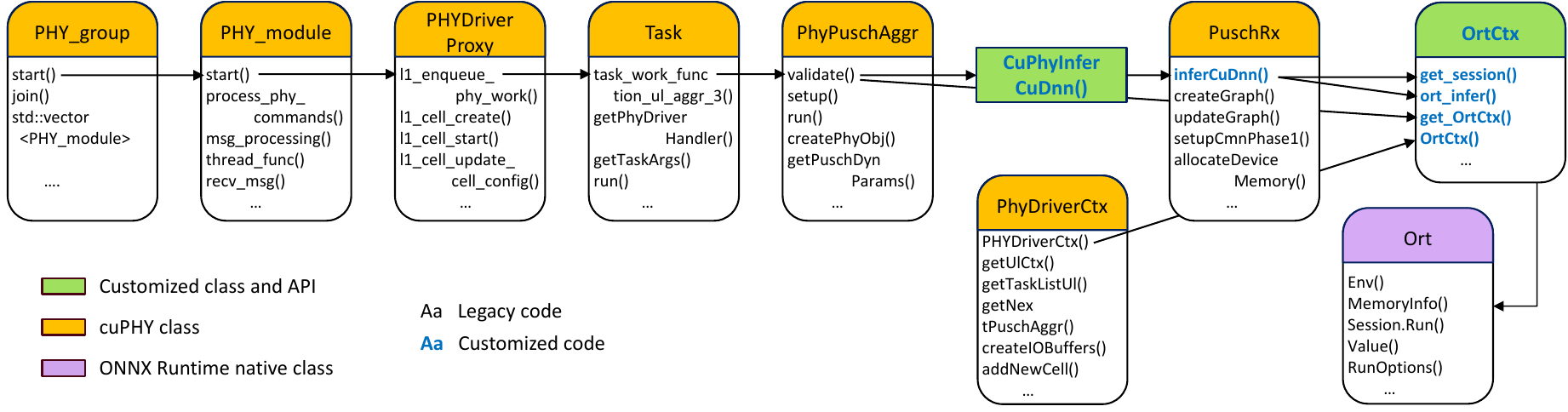}
    \setlength\belowcaptionskip{-.3cm} 
        \setlength\abovecaptionskip{0cm}
  \caption{\justifying The workflow diagram for the OrtCtx class for performing inference.}
  \label{uml}
\end{figure*}

\textbf{\dapp System Structure.}
As the core component of \dapp, the inference module functions as a post-processing unit following the execution of the \gls{pusch} pipeline, ensuring that ongoing \gls{ul}/\gls{dl} pipeline processes remain unaffected, as illustrated in Fig.~\ref{cuda_graph}. This functionality is efficiently implemented through the function \texttt{CuPhyInferCuDnn()}---a customized \gls{api} hosted by \gls{cuphy} and managed by the \gls{cuphy} driver, as indicated in Fig.~\ref{uml}. 
This \gls{api} further employs the \texttt{Session.Run()} method from the \gls{ort} framework to execute the actual inference process. To prevent \dapp from interfering with the tightly time-coupled operations of the \gls{pusch} or other channels, the \gls{cuphy} driver delegates the processing to a newly instantiated and independent \gls{cpu} thread. 
This thread is assigned to a dedicated \gls{cpu} core---isolated from other operations---to improve stability. 

\textbf{ORT Integration and Tuning.}
\gls{ort} is built to execute models in the \gls{onnx} format, which is an open standard for representing \gls{ml} models. \gls{ai} operations using \gls{ort} can be executed on either the \gls{cpu}, utilizing the \gls{cpu} execution provider, or on the \gls{gpu}, utilizing \gls{cuda} or \gls{tensorrt} execution providers. Our implementation supports all three options, with processing latency shown in Tab.~\ref{tab:orttimes}, for inference using standalone \gls{ort} (i.e., independently of the dApp implementation) and the model discussed in Sec.~\ref{subsec:cnn_model}. \gls{cpu}-based inference takes approximately $30.354$~ms for a 24-core Intel Xeon Gold CPU, while \gls{gpu}-based inference using \gls{tensorrt} and \gls{cuda} takes $0.524$~ms and $3.675$~ms using an NVIDIA A100, respectively. 

\begin{table}[t]
    \centering
    \captionsetup{aboveskip=.1pt, belowskip=.1pt} 
    \caption{\justifying Inference time between different execution providers for standalone ORT.}
    \label{tab:orttimes}
    \footnotesize
    \begin{tabular}{p{0.35\linewidth} p{0.15\linewidth} p{0.15\linewidth} p{0.15\linewidth}}  
        \toprule
       \textbf{ORT Execution Provider} & CPU & CUDA & TensorRT \\
        \midrule
        \textbf{Inference Time (ms)} & 30.354 & 3.675 & 0.524 \\
        \bottomrule
    \end{tabular}
\end{table} 

Although both \gls{cuda} and \gls{tensorrt} utilize the \gls{gpu} for inference tasks, \gls{tensorrt} is specifically optimized for high performance. NVIDIA \gls{tensorrt} is an \gls{sdk} for deep learning model optimization, featuring an inference optimization engine and runtime environment. As evidenced in Tab.~\ref{tab:orttimes}, it achieves lower execution times than both \gls{cpu} and \gls{cuda} providers. The framework parses the \gls{onnx} model, applies optimizations such as layer fusion, and selects efficient \gls{cuda} kernels by leveraging libraries like \gls{cudnn} and \gls{cublas}. Consequently, \gls{tensorrt} is used exclusively in our experiments to maximize performance. 

Based on our measurements, however, the \gls{gpu}-based \gls{ort} mode comes with a significant delay during initial execution---$1.25$\:s and $9.65$\:s with \gls{cuda} and \gls{tensorrt}, respectively---for the model described in Sec.~\ref{subsec:cnn_model}. 
These findings identify the root cause of the timing issue as the long \gls{gpu} warm-up phase.
This is because \gls{ort} requires several seconds to initialize and allocate \gls{gpu} resources essential for inference. In particular, during the first call to \texttt{Session.Run()}, \gls{ort} performs several critical setup steps:
(i) Memory allocation for the model tensors;
(ii) Graph optimization and compilation to ensure efficient runtime execution;
(iii) Execution provider setup, such as configuring \gls{cuda} for \gls{gpu} inference; and
(iv) Caching mechanisms for future executions.
This warm-up latency conflicts with the strict timing constraints of the \gls{ul} slot in \gls{5g} \gls{nr} systems, making it a challenge for real-time execution. However, once initialized, these setup steps are cached, allowing subsequent invocations to bypass the preparatory steps and run significantly faster.

\textbf{PHY Layer and dApp Setup.} Therefore, we design the \dapp dApp so that it performs an initial warm-up inference before the base station becomes operational. 
Particularly, during the initialization of \gls{cubb}, a preliminary inference step is performed on dummy data encompassing all input features. Then a pointer to the \gls{ort} run session (already initialized) is passed to \dapp. 
To support this, \dapp is encapsulated within a customized \gls{ort} context class, \texttt{OrtCtx}, which is instantiated within the \gls{cuphy} driver context class, \texttt{PhyDriverCtx}, as illustrated in Fig.~\ref{uml}.
This manages instances of transmit and receive processing pipelines for \gls{ul} and \gls{dl} slots, as well as individual classes corresponding to each \gls{phy} channel, among others. Additionally, it handles the allocation of \gls{gpu} resources and manages the \gls{gpu} footprint for all \gls{ul} and \gls{dl} executions. \texttt{PhyDriverCtx} is instantiated as the whole physical layer instance starts, making it possible to execute the warm-up phase for \gls{ort}. 

As part of this process, the \texttt{OrtCtx} class constructor initializes and configures the dynamically allocated instance of an \gls{ort} inference session as a smart pointer, using the native \gls{ort} \gls{api} and leveraging its comprehensive classes and methods, as shown in Fig.~\ref{uml}.  
The process starts with creating an \texttt{Ort::Env()} object for logging and runtime management, followed by configuring session options for optimization, including graph processing, threading, memory allocation, and \gls{gpu}-specific settings for \gls{cuda} and \gls{tensorrt}.
The dummy input data for warm-up inference is allocated in \gls{cpu} memory and transferred to the \gls{gpu} for execution, encapsulated as \texttt{Ort::Value()} tensors. \texttt{Session.Run()} processes the data through the pre-trained \gls{onnx} model and generates output tensors containing class probabilities, selecting the highest probability as the final prediction. 

\textbf{Runtime Inference.} Upon initialization, the \texttt{OrtCtx} object is passed to the \gls{pusch} channel via the physical \gls{pusch} aggregate class, \texttt{PhyPuschAggr}, and invoked during the validation phase as a callable function, \texttt{get\_OrtCtx()}. This validation phase is a post-processing step always executed after the \gls{pusch} pipeline. Inference in the \gls{pusch} pipeline is orchestrated by \texttt{inferCuDnn()}, a member of \texttt{PuschRx} class, and invoked by \texttt{CuPhyInferCuDnn()} \gls{api}, as illustrated in Fig.~\ref{uml}. It retrieves the inference session from the \texttt{OrtCtx} object via \texttt{get\_session()}, reusing the session configuration and runtime environment. We encapsulate input features into \gls{onnx}-compatible tensors using \texttt{ort\_infer()}, using \gls{gpu} memory for \gls{cuda}/\gls{tensorrt} or \gls{cpu} memory otherwise. When performing inference on the \gls{gpu}, the input features are transferred directly within \gls{gpu} memory without overhead. Tensors follow predefined shapes and types, with input/output names built dynamically for runtime flexibility.

\textbf{dApp-based Automated Data Collection.} We also designed the \dapp dApp to perform automated data collection to create datasets for offline model training. This leverages a \texttt{CuPhyInfer\-CuDnn()} \gls{api} to stream data into an \gls{hdf5} file at a frequency of one every 10~\gls{ul} slots. Different from what happens for online inference, the \gls{api} asynchronously transfers features to be stored from \gls{gpu} to \gls{cpu} and writes to disk thereafter. 
This avoids many \gls{gpu} operations, including
data conversion to tensors and subsequent \gls{gpu} memory deallocation. 
Besides, to avoid interference with the tightly coupled operations of \gls{pusch} or other channels, the \gls{cuphy} driver delegates data logging to a newly instantiated \gls{cpu} thread operating in detached mode, ensuring uninterrupted processing.

\vspace{-10pt}
\subsection{AI-Based Interference Detection}
\label{subsec:cnn_model}

\dapp leverages a \gls{cnn} for real-time interference classification, using a data-driven methodology that allows the network to identify complex patterns in incoming \gls{ul} data. 

\textbf{Input Features.}
As discussed above, the input features include \gls{iq} samples for the \gls{pusch}, as well as additional \glspl{kpm} representing \gls{rssi}, \gls{rsrp}, \gls{sinr}, \gls{mcs} index, \gls{mcs} table index, total count of \gls{cb} errors, and number of \glspl{cb} derived from \gls{cb}-\glspl{crc}. As part of the dApp design, one challenge was ensuring the compatibility of heterogeneous numerical formats, such as \textit{float16} (for \gls{iq}), \textit{float32} (for \gls{rssi}, \gls{rsrp}, \gls{sinr}), and \textit{int32} (for \gls{mcs} and \gls{cb} inputs), across frameworks such as \gls{tf} (offline training) and \gls{cubb}.

Unlike traditional approaches that rely on heatmaps of \gls{iq} samples for interference detection~\cite{8885870,9600524}, \dapp processes raw \gls{iq} data directly and incorporates additional features to improve classification accuracy.
\gls{iq} samples are extracted prior to the \gls{mmse}-\gls{irc} equalizer, which is designed to mitigate channel distortion and interference. This pre-equalization extraction provides an unaltered view of interference effects, allowing for more accurate analysis. 
In \gls{cubb}, the \gls{pusch} channel \gls{iq} samples are stored as 32-bit words (with two 16 bits floats for each component), in blocks of $14\times273\times12$ contiguous words, where $14$ is the number of symbols per slot, $273$ is the maximum number of \glspl{prb} with $12$ subcarriers each.  
\gls{ort} converts the \gls{iq} samples into a matrix with dimensions $14, 3276, 2$, to align with \gls{tf}'s expected format for inference, interpreting the third dimension as the real and imaginary components of the \glspl{iq}.

Additionally, specific input features are transformed to ensure data compatibility. In the native \gls{cb}-\gls{crc} \gls{cuda} kernel, the \gls{cb} errors for each \gls{pusch} channel are output as an array, where each non-zero value represents a corrupted \gls{cb} within the \gls{tb}, resulting in a dynamic size. To meet the fixed-size input requirements of the learning model, two additional variables are added to the \gls{gpu}: one to determine the number of corrupted \glspl{cb} in a \gls{tb} and another to find the total number of \glspl{cb}. These operations are seamlessly integrated into the kernel 11 in Fig.~\ref{cuda_graph}.

\begin{figure}[htbp]
  \centering
  \includegraphics[width=0.9835\columnwidth]{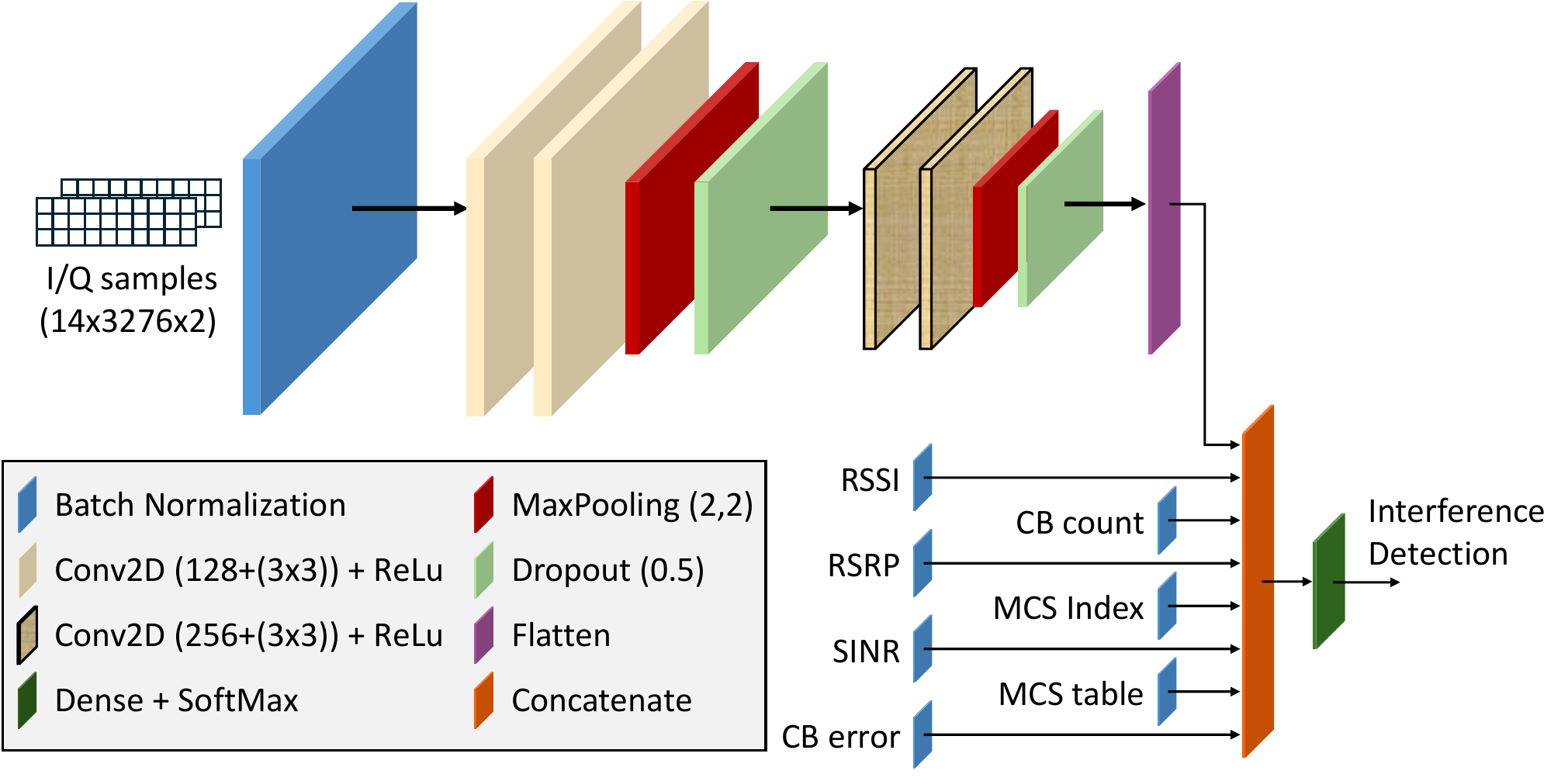}
  \setlength\abovecaptionskip{2pt}
  \setlength\belowcaptionskip{-.2cm} 
  \caption{\justifying \dapp's CNN architecture showing inputs (I/Q and scalar features), layer specifications, and the binary interference detection output.}
  \label{fig:cnn}
\end{figure}

\textbf{CNN Architecture.}
We choose CNNs for their ability to extract complex patterns from OFDM-based IQ samples and their consistent outperformance of alternatives like LSTMs and ResNets. Fig.~\ref{fig:cnn} illustrates the selected architecture. The network features seven layers. Two convolutional blocks perform feature extraction on \gls{pusch} \gls{iq} samples, each containing two Conv2D layers (128 and 256 filters, $3\times3$ kernel, ReLU activation) and one $2\times2$ MaxPooling layer. Then, the extracted features are flattened and concatenated with the additional normalized structured inputs (i.e., \gls{rssi}, \gls{rsrp}, \gls{sinr}, \gls{mcs} index, \gls{mcs} table index, total count of \gls{cb} errors, and number of \glspl{cb}). This final representation is processed through a fully connected Dense layer with softmax activation, which performs the final interference detection.
The starting point of our proposed \gls{cnn} is the VGG16 model, a widely recognized architecture in image processing. We systematically reduced its depth and identified a two-block variant that provided the best trade-off between computational efficiency and classification performance.

To enhance generalization and mitigate overfitting, we integrate dropout layers and an L2 regularizer. Dropout layers randomly deactivate neurons during training, encouraging the model to learn underlying patterns rather than memorizing the training data, while the L2 regularizer penalizes large weights, promoting simpler, more generalized models. Additionally, to address class imbalances (e.g., in case of data collections with different number of samples for different radios, or interference conditions), we use a weighted loss function instead of a standard one, which tends to favor the majority class. By assigning a greater weight to the minority class, the loss function ensures that its contributions have a stronger influence on the overall loss computation.

\textbf{Transfer Learning.}
We employ \gls{tl} using heterogeneous datasets from diverse \gls{rf} environments, detailed in Sec.~\ref{subsec:data}, to enhance model generalization capabilities. \gls{tl} is a machine learning technique wherein knowledge acquired from a source domain is leveraged to improve performance on a target domain through model adaptation. In our implementation, we partially fine-tune the foundational \gls{cnn} model (Fig.~\ref{fig:cnn}), trained on data from multiple deployment location (see Fig.~\ref{fig:map_EXP}), and then tune to specific deployments by freezing the first block and fine-tuning the remaining layers with site-specific data. Thus, we allow the model to learn from a broader range of conditions and fine-tune it for a specific deployment, retaining knowledge of diverse scenarios by preserving foundational feature representations and enhancing performance.

\section{Data and Evaluation Framework}
\label{sec:framework}

This section presents the setup for the training, testing, and evaluation of \dapp, from the generation of synthetic data with simulations to experimental \gls{ota} data collection and evaluation. We also discuss preprocessing and offline training procedures.

\vspace{-5pt}
\subsection{Primer on X5G}
\label{subsec:x5g}


Real-world data collection, experiments, and validation of \dapp are performed using the X5G platform.
X5G is an O-RAN-compliant, multi-vendor private \gls{5g} network featuring a multi-node deployment of the NVIDIA Aerial framework~\cite{arc-ota}, spanning the entire floor of the EXP building on the Boston, MA, campus of Northeastern University~\cite{villa2024x5g}.
The platform is fully open and programmable, allowing researchers to modify any part of it for experimentation and testing, including the integration and development of dApps like \dapp, within a testbed offering production-ready performance and capabilities.
Each node features \gls{oai} for the upper layers of the protocol stack (\gls{cu} and \gls{du}-High) and the NVIDIA Aerial \gls{sdk} for the \gls{du}-Low. A structured networking infrastructure, consisting of various switches and connections, enables switching between \glspl{ru} from different vendors, including Foxconn operating in the n78 band. 
Additionally, the \gls{cn} is sourced from various open projects, such as Open5GS. X5G supports a range of \gls{cots} \glspl{ue}, including OnePlus, iPhone, and Samsung, as well as \gls{5g} modules like Sierra boards. Finally, it integrates the \gls{osc} \gls{near-rt-ric} for xApp development.
All X5G \gls{ran} software runs inside Docker containers on dedicated \gls{ran} servers, including Gigabyte E251-U70 machines. Each Gigabyte server is equipped with a 24-core Intel Xeon Gold \gls{cpu}, $96$~GB of RAM, and a Broadcom \gls{pci} switch with two slots, hosting an NVIDIA A100 \gls{gpu} and a Mellanox ConnectX-6 Dx \glspl{nic}. This setup enables direct connectivity, bypassing \gls{cpu} involvement, while ensuring high transfer speeds and providing the necessary \gls{gpu} computational power to run the \gls{ran} software along with additional workloads, such as \dapp.

\vspace{-5pt}
\subsection{Empirical \gls{ota} Data Collection, Automated Labeling, and Preprocessing}
\label{subsec:data}

\begin{figure}[htbp]
  \centering
  \includegraphics[width=\columnwidth]{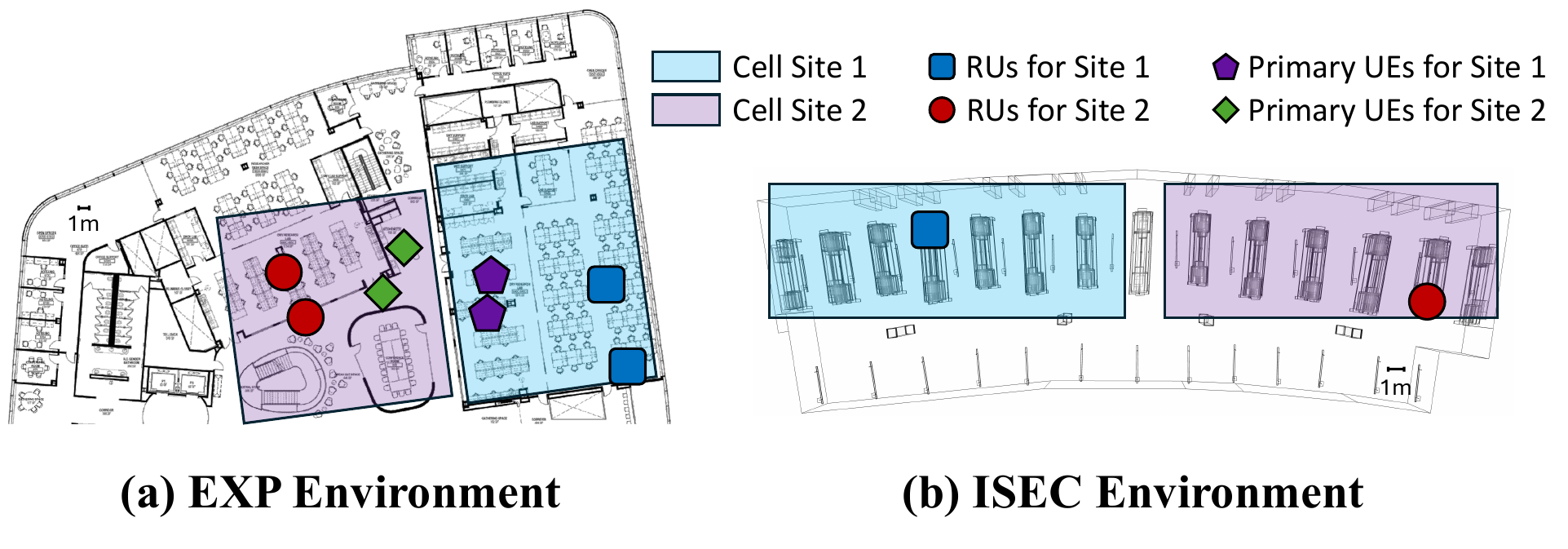}
  
  \setlength\abovecaptionskip{0cm} 
  \caption{\justifying Indoor experiment layout maps of Northeastern University EXP (a) and ISEC (b) buildings, showing: the designated regions for the two cell sites, where the corresponding UEs can be located; RU locations for each respective cell site; and primary UE locations, where most of the data is collected.}
  \label{fig:map_EXP}
\end{figure}

All experiments involving interference are conducted in a controlled indoor environment within the Northeastern University EXP building in Boston, MA, as depicted in the indoor experiment map shown in Fig.~\ref{fig:map_EXP}a. The layout includes: two cell sites (shown in light blue and pink regions), where the corresponding \glspl{ue} can be located; the \gls{ru} locations for each cell site, with only one active at a time per site (represented by blue and red icons); and the two primary \gls{ue} locations for the two cell sites, where most of the data is collected (violet and green icons).
Each experiment involves both cell sites and one \gls{ue} per cell at a time, tested under \gls{los} and \gls{nlos} channel conditions.
Similarly, we collected additional data in a second indoor environment within the Northeastern University ISEC building in Boston, MA, (Fig.~\ref{fig:map_EXP}b) as mentioned in Sec.~\ref{subsec:cnn_model}.

During each data collection session, the \gls{ue} connects to its \gls{ru} and generates \gls{ul} traffic using iPerf.
We categorize the \gls{ul} traffic levels at the \glspl{gnb} into two distinct classes---High Traffic and No Traffic---based on the \gls{tb} size, to improve interference identification accuracy. In the No Traffic scenario, the \gls{ue} remains connected to the \gls{gnb} while exchanging minimal control information necessary for maintaining the connection stability. In this case, we observe that the \gls{tb} size is typically around $185$~bytes and does not exceed $1056$~bytes, with the number of \glspl{cb} limited to one. 
The \glspl{cb} from a \gls{tb} are segmented and \gls{ldpc}-encoded using base graphs defined by the 3GPP for error correction. \gls{ldpc} base graph 2 is used for smaller \glspl{cb}, with each \gls{cb} sized at $480$~bytes, while base graph 1 is used for larger ones, with each \gls{cb} sized at $1056$~bytes. 
In contrast, High Traffic \gls{ul} transmissions produce sufficiently strong signals that can interfere with neighboring cells. In this scenario, the \gls{tb} size typically exceeds $1056$~bytes, and the number of \glspl{cb} is greater than one.

As depicted in Fig.~\ref{combined_plots}, there is a strong correlation between the \gls{tb} size and the number of \glspl{cb} (Fig.~\ref{fig:cb_tbsize}), as well as between the \gls{ul} throughput and the \gls{cb} count (Fig.~\ref{fig:cb_ulthru}), reinforcing the decision to use the \gls{cb} count as the primary criterion for traffic classification.
Building upon this, we create a systematic labeling methodology based on the \gls{ul} traffic levels to classify the \gls{ul} transmissions as either affected by interference ('INTERF') or unaffected ('CLEAN').
The labeling rules, shown in Tab.~\ref{tab:lookup}, are defined as follows:
\begin{itemize}[noitemsep, topsep=2pt, leftmargin=12pt]
    \item No Traffic/High Traffic: if one \gls{ue} transmits with 'High Traffic' while the other has 'No Traffic', the 'High Traffic' time windows on one side are used to label the other side as being affected by interference, designated as 'INTERF', while the transmitting side is labeled as ‘CLEAN’.
    \item High Traffic on both sides: if both \glspl{ue} transmit simultaneously to different \glspl{ru} with 'High Traffic', the corresponding samples are labeled as 'INTERF'.
    \item No Traffic on both sides: if both \glspl{ue} are transmitting with 'No Traffic', the samples are labeled as 'CLEAN', i.e., no interference.
    \item Single-active Transmission: if only one \gls{ue} transmits while the other remains in airplane mode (i.e., not connected to the base station), the samples are labeled as ‘CLEAN’.
\end{itemize}
A threshold of 1 is set for the \gls{cb} count in an \gls{ul} slot to identify high-traffic windows on one \gls{gnb}, which are subsequently used to mark instances of interference on the other \gls{gnb}.
Notably, the exact thresholding mechanism is utilized to evaluate inference performance in \gls{ota} data scenarios by identifying high-traffic time windows, which are then used to assign ground truth labels of ‘CLEAN’ or ‘INTERF’ to each input example, thereby enabling the effective monitoring of spectrum sensing capabilities.

\begin{figure}[t]
    \centering
    \begin{subfigure}{0.465\textwidth} 
        \centering
    \setlength\fwidth{\linewidth}
    \setlength\fheight{.29\linewidth}{\input{figures_tex/tbsize_cb.tex}}
        \setlength\abovecaptionskip{0cm}
        \setlength\belowcaptionskip{-.1cm}
        \caption{\centering TB size in bytes (scaled down by 1000) versus CB count}
        \label{fig:cb_tbsize}
    \end{subfigure}
    \hfill
        \begin{subfigure}{0.465\textwidth} 
        \centering
    \setlength\fwidth{\linewidth}
    \setlength\fheight{.29\linewidth}
    \input{figures_tex/ul_throughput_vs_cb_count.tex}
        \setlength\abovecaptionskip{-.4cm}
        \setlength\belowcaptionskip{0cm}
        \caption{\centering UL throughput versus CB count}
        \label{fig:cb_ulthru}
    \end{subfigure}

    \captionsetup{aboveskip=.1pt, belowskip=.1pt} 
    \caption{\justifying UL throughput, TB size, and CB count across different scenarios, demonstrating their close correlation.}
    \label{combined_plots}
\end{figure}

To efficiently manage \gls{gpu} memory during offline training---particularly given the relatively large input feature consisting of 3D \gls{iq} samples of size $14\times3276\times2$---we use \gls{tf}’s Sequence class to implement a custom data generator. The data generator dynamically loads small batches into \gls{gpu} memory, rather than loading the entire dataset at once, thereby optimizing memory usage and avoiding excessive memory strain.

\begin{table}[t]
    \vspace{-10pt}
    \centering
    \captionsetup{aboveskip=.1pt, belowskip=.1pt} 
    \caption{\justifying Lookup Table for gNB UL Traffic and Labels}
    \label{tab:lookup}
    \footnotesize
    \begin{tabular}{p{0.2\linewidth} p{0.2\linewidth} p{0.2\linewidth} p{0.2\linewidth}}  
        \toprule
       \textbf{gNB1 Traffic} & \textbf{gNB2 Traffic} & \textbf{gNB1 UL Label} & \textbf{gNB2 UL Label} \\
        \midrule
        High Traffic & No Traffic & CLEAN & INTERF \\
        No Traffic & High Traffic & INTERF & CLEAN \\
        High Traffic & High Traffic & INTERF & INTERF \\
        No Traffic & No Traffic & CLEAN & CLEAN \\
        No UE & No/High Traffic & NA & CLEAN \\
        No/High Traffic & No UE & CLEAN & NA \\
        \bottomrule
    \end{tabular}
\end{table}

\subsection{Synthetic Data}
\label{sec:synth_data}
We also perform an extensive data collection campaign using the NVIDIA Aerial simulator nr\_sim, which complements the \gls{cubb} framework by extending the 5G MATLAB Toolbox. Through this, we collect data and evaluate the model in a larger variety of configurations compared to those supported in our experimental setup. The parameters that we sweep include \gls{snr}, \gls{sir}, \gls{mcs} index, channel type, delay profile, interference channel delay profile, numerology, and carrier frequency. In addition, we simulate up to 5 interfering \glspl{ue}, so that the classification can be extended to include additional \glspl{ue}.

To simulate interference, an in-band interference signal is artificially generated by transmitting \gls{ofdm} signals from each antenna independently through a \gls{tdl} channel characterized by high delay spread and Doppler shift. These conditions emulate a dynamic and realistic environment affected by multipath propagation and high mobility, thus creating a robust testbed for interference mitigation. The processed signal is then multiplied by a modified all-zero matrix with randomly distributed contiguous ones, introducing localized and burst interference in the \glspl{re} of the \gls{ofdm} slot. This process ensures random patterns for contiguous ones across antenna streams while maintaining the average power of the interference signal. The resulting combination of randomized interfering signals, thermal noise, and received \gls{ofdm} signal at the \gls{gnb} closely mimics real-world interference effects.

\section{Performance Evaluation and Experimental\\ Results}
\label{sec:results}

This section presents experimental results to benchmark the performance of \dapp across various model configurations, using the simulation setup described in Sec.~\ref{sec:synth_data} as well as over \gls{ota} tests.

\begin{figure}[htbp]
    \centering
    \setlength\fwidth{1\linewidth}
    \setlength\fheight{.3\linewidth}
\begin{tikzpicture}

\definecolor{black38}{RGB}{38,38,38}
\definecolor{gray176}{RGB}{176,176,176}

\definecolor{color0}{RGB}{230,240,255}    
\definecolor{color1}{RGB}{198,219,239}    
\definecolor{color2}{RGB}{158,202,225}    
\definecolor{color3}{RGB}{107,174,214}    
\definecolor{color4}{RGB}{66,146,198}     
\definecolor{color5}{RGB}{33,113,181}     
\definecolor{color6}{RGB}{8,81,156}       
\definecolor{color7}{RGB}{0,55,120}       

\begin{axis}[
    width=.937\fwidth,
    height=1.5\fheight,
    tick align=inside,
    tick pos=left,
    x grid style={gray176},
    xlabel={Predicted label},
    xlabel style={font=\footnotesize},
    xmin=0, xmax=6,
    xtick style={color=black},
    xtick={0.5,1.5,2.5,3.5,4.5,5.5},
    xticklabels={0,1,2,3,4,5},
    y dir=reverse,
    y grid style={gray176},
    ylabel={True label},
    ylabel style={font=\footnotesize},
    ymin=0, ymax=6,
    ytick style={color=black},
    ytick={0.5,1.5,2.5,3.5,4.5,5.5},
    yticklabel style={rotate=90.0},
    yticklabels={0,1,2,3,4,5},
    xticklabel style={font=\footnotesize},
    yticklabel style={font=\footnotesize},
    colormap={bluegrad}{
        rgb255(0pt)=(230,240,255);
        rgb255(1pt)=(230,240,255);
        rgb255(3pt)=(198,219,239);
        rgb255(10pt)=(158,202,225);
        rgb255(30pt)=(107,174,214);
        rgb255(60pt)=(66,146,198);
        rgb255(90pt)=(33,113,181);
        rgb255(95pt)=(8,81,156);
        rgb255(100pt)=(0,55,120)
    },
    colorbar,
    colorbar style={
        title={Accuracy (\%)},
        title style={font=\footnotesize, rotate=90, yshift=-10mm, xshift=-12.5mm},
        ylabel style={font=\footnotesize},
        yticklabel style={font=\footnotesize},
        ytick={0,20,40,60,80,100},
        yticklabels={0,20,40,60,80,100},
        width=2mm,
        yticklabel pos=right
    },
    point meta min=0,
    point meta max=100
]


\fill[color7] (0,0) rectangle (1,1); 
\fill[color1] (1,0) rectangle (2,1); 
\fill[color0] (2,0) rectangle (3,1); 
\fill[color0] (3,0) rectangle (4,1); 
\fill[color0] (4,0) rectangle (5,1); 
\fill[color0] (5,0) rectangle (6,1); 

\fill[color2] (0,1) rectangle (1,2); 
\fill[color6] (1,1) rectangle (2,2); 
\fill[color1] (2,1) rectangle (3,2); 
\fill[color0] (3,1) rectangle (4,2); 
\fill[color0] (4,1) rectangle (5,2); 
\fill[color0] (5,1) rectangle (6,2); 

\fill[color0] (0,2) rectangle (1,3); 
\fill[color0] (1,2) rectangle (2,3); 
\fill[color7] (2,2) rectangle (3,3); 
\fill[color0] (3,2) rectangle (4,3); 
\fill[color0] (4,2) rectangle (5,3); 
\fill[color0] (5,2) rectangle (6,3); 

\fill[color0] (0,3) rectangle (1,4); 
\fill[color0] (1,3) rectangle (2,4); 
\fill[color0] (2,3) rectangle (3,4); 
\fill[color7] (3,3) rectangle (4,4); 
\fill[color0] (4,3) rectangle (5,4); 
\fill[color0] (5,3) rectangle (6,4); 

\fill[color0] (0,4) rectangle (1,5); 
\fill[color0] (1,4) rectangle (2,5); 
\fill[color0] (2,4) rectangle (3,5); 
\fill[color0] (3,4) rectangle (4,5); 
\fill[color7] (4,4) rectangle (5,5); 
\fill[color0] (5,4) rectangle (6,5); 

\fill[color0] (0,5) rectangle (1,6); 
\fill[color0] (1,5) rectangle (2,6); 
\fill[color0] (2,5) rectangle (3,6); 
\fill[color0] (3,5) rectangle (4,6); 
\fill[color0] (4,5) rectangle (5,6); 
\fill[color7] (5,5) rectangle (6,6); 

\draw (axis cs:0.5,0.5) node[scale=1.2, text=white, rotate=0.0, align=center]{\footnotesize 96.93};
\draw (axis cs:1.5,0.5) node[scale=1.2, text=black38, rotate=0.0, align=center]{\footnotesize 2.66};
\draw (axis cs:2.5,0.5) node[scale=1.2, text=black38, rotate=0.0, align=center]{\footnotesize 0.41};
\draw (axis cs:3.5,0.5) node[scale=1.2, text=black38, rotate=0.0, align=center]{\footnotesize 0.00};
\draw (axis cs:4.5,0.5) node[scale=1.2, text=black38, rotate=0.0, align=center]{\footnotesize 0.00};
\draw (axis cs:5.5,0.5) node[scale=1.2, text=black38, rotate=0.0, align=center]{\footnotesize 0.00};

\draw (axis cs:0.5,1.5) node[scale=1.2, text=black38, rotate=0.0, align=center]{\footnotesize 4.02};
\draw (axis cs:1.5,1.5) node[scale=1.2, text=white, rotate=0.0, align=center]{\footnotesize 93.87};
\draw (axis cs:2.5,1.5) node[scale=1.2, text=black38, rotate=0.0, align=center]{\footnotesize 2.11};
\draw (axis cs:3.5,1.5) node[scale=1.2, text=black38, rotate=0.0, align=center]{\footnotesize 0.00};
\draw (axis cs:4.5,1.5) node[scale=1.2, text=black38, rotate=0.0, align=center]{\footnotesize 0.00};
\draw (axis cs:5.5,1.5) node[scale=1.2, text=black38, rotate=0.0, align=center]{\footnotesize 0.00};

\draw (axis cs:0.5,2.5) node[scale=1.2, text=black38, rotate=0.0, align=center]{\footnotesize 0.00};
\draw (axis cs:1.5,2.5) node[scale=1.2, text=black38, rotate=0.0, align=center]{\footnotesize 0.00};
\draw (axis cs:2.5,2.5) node[scale=1.2, text=white, rotate=0.0, align=center]{\footnotesize 99.88};
\draw (axis cs:3.5,2.5) node[scale=1.2, text=black38, rotate=0.0, align=center]{\footnotesize 0.12};
\draw (axis cs:4.5,2.5) node[scale=1.2, text=black38, rotate=0.0, align=center]{\footnotesize 0.00};
\draw (axis cs:5.5,2.5) node[scale=1.2, text=black38, rotate=0.0, align=center]{\footnotesize 0.00};

\draw (axis cs:0.5,3.5) node[scale=1.2, text=black38, rotate=0.0, align=center]{\footnotesize 0.00};
\draw (axis cs:1.5,3.5) node[scale=1.2, text=black38, rotate=0.0, align=center]{\footnotesize 0.00};
\draw (axis cs:2.5,3.5) node[scale=1.2, text=black38, rotate=0.0, align=center]{\footnotesize 0.00};
\draw (axis cs:3.5,3.5) node[scale=1.2, text=white, rotate=0.0, align=center]{\footnotesize 100.00};
\draw (axis cs:4.5,3.5) node[scale=1.2, text=black38, rotate=0.0, align=center]{\footnotesize 0.00};
\draw (axis cs:5.5,3.5) node[scale=1.2, text=black38, rotate=0.0, align=center]{\footnotesize 0.00};

\draw (axis cs:0.5,4.5) node[scale=1.2, text=black38, rotate=0.0, align=center]{\footnotesize 0.00};
\draw (axis cs:1.5,4.5) node[scale=1.2, text=black38, rotate=0.0, align=center]{\footnotesize 0.00};
\draw (axis cs:2.5,4.5) node[scale=1.2, text=black38, rotate=0.0, align=center]{\footnotesize 0.00};
\draw (axis cs:3.5,4.5) node[scale=1.2, text=black38, rotate=0.0, align=center]{\footnotesize 0.00};
\draw (axis cs:4.5,4.5) node[scale=1.2, text=white, rotate=0.0, align=center]{\footnotesize 100.00};
\draw (axis cs:5.5,4.5) node[scale=1.2, text=black38, rotate=0.0, align=center]{\footnotesize 0.00};

\draw (axis cs:0.5,5.5) node[scale=1.2, text=black38, rotate=0.0, align=center]{\footnotesize 0.00};
\draw (axis cs:1.5,5.5) node[scale=1.2, text=black38, rotate=0.0, align=center]{\footnotesize 0.00};
\draw (axis cs:2.5,5.5) node[scale=1.2, text=black38, rotate=0.0, align=center]{\footnotesize 0.00};
\draw (axis cs:3.5,5.5) node[scale=1.2, text=black38, rotate=0.0, align=center]{\footnotesize 0.00};
\draw (axis cs:4.5,5.5) node[scale=1.2, text=black38, rotate=0.0, align=center]{\footnotesize 0.00};
\draw (axis cs:5.5,5.5) node[scale=1.2, text=white, rotate=0.0, align=center]{\footnotesize 100.00};

\textbf{}\end{axis}

\end{tikzpicture}
        \captionsetup{aboveskip=.1pt, belowskip=.1pt} 
    \caption{\justifying Confusion matrix of the model tested on synthetic data with 13800 samples per class.}
    \label{fig:conf_matrix_nrsim}
    \Description{Confusion matrix of the selected model tested on synthetic data with 13800 samples per class.}
\end{figure}
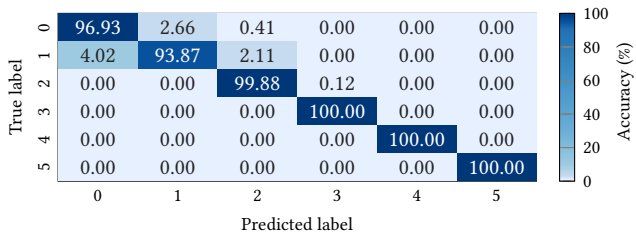

We test the model, shown in Fig.~\ref{fig:cnn}, on synthetic data with scenarios involving 0 to 5 interferers, totaling $13800$~samples per class, to ensure consistent performance across different parameters described in Sec.~\ref{sec:synth_data} before \gls{ota} deployment in \dapp. The resulting confusion matrix in Fig.~\ref{fig:conf_matrix_nrsim} indicates that the model achieves high accuracy (above 96\%) for scenarios with no interferer and 2 or more interferers. For scenarios with a single interferer, accuracy remains solid at $93.87$\%, with the slight drop likely due to the lower impact of a single interferer on the signal. Thus, the model becomes a strong candidate for \gls{ota} deployment, along with several other models with minor modifications explained in Sec~\ref{sec:ota_results}.

\vspace{-10pt}
\subsection{OTA Evaluation}
\label{sec:ota_results}

To select the best generalized model suitable for diverse \gls{rf} environments, we conduct a series of \gls{ota} experiments evaluating eight configurations of the underlying \gls{cnn} architecture shown in Fig.~\ref{fig:cnn}.
We represent each model as \{[$\alpha$,$\beta$],$\gamma$\}, where $\alpha$ and $\beta$ denote the number of filters in the convolutional layers of the first and second blocks, and $\gamma$ represents the batch size.
We select two configurations for [$\alpha$, $\beta$]: [64, 128] and [128, 256], inspired by VGG16's first blocks for a lightweight model, with $\gamma$ ranging over 16, 32, 64, and 128.

We carry out experiments at various \gls{ru} locations and \gls{ue} positions across the map, as shown in Fig.~\ref{fig:map_EXP}.
For fair comparability, we evaluate different models using the same dataset with data logging. The data collected from one side is labeled with ground truth based on the traffic level on the other side, as detailed in Tab.~\ref{tab:lookup}.
We then compute metrics such as accuracy, measuring the ratio of correctly predicted samples to the total number of samples, specificity, evaluating the identification of 'CLEAN' examples, and recall, assessing the detection of 'INTERF' examples.

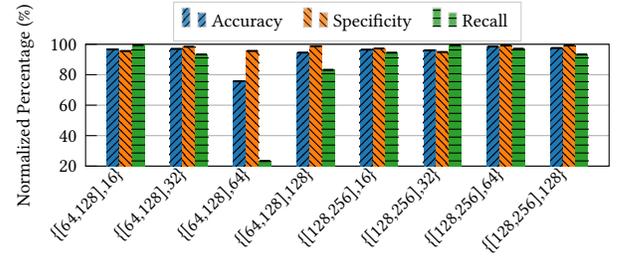
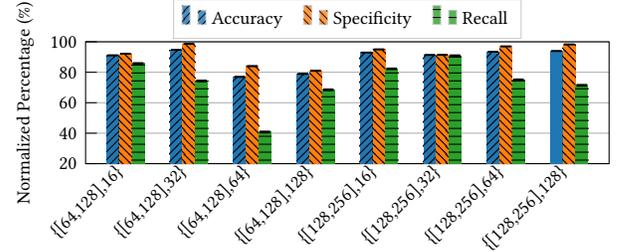
\begin{figure}[t]
    \centering
    \begin{subfigure}{0.48\textwidth} 
            \centering
        \setlength\fwidth{\linewidth}
        \setlength\fheight{.15\linewidth}
\begin{tikzpicture}

\definecolor{gray176}{RGB}{176,176,176}
\definecolor{gray31119180}{RGB}{31,119,180}
\definecolor{green4416044}{RGB}{44,160,44}
\definecolor{orange25512714}{RGB}{255,127,14}
\definecolor{white204}{RGB}{204,204,204}

\definecolor{gray176}{RGB}{176,176,176}
\definecolor{blue31119180}{RGB}{31,119,180}
\definecolor{green4416044}{RGB}{44,160,44}
\definecolor{orange25512714}{RGB}{255,127,14}
\definecolor{white204}{RGB}{204,204,204}

\begin{axis}[
width=\fwidth,
height=2.5\fheight,
legend cell align={left},
legend style={
  at={(.5,1.35)},
  anchor=north,
  draw=white204,
  fill opacity=0.8,
  draw opacity=1,
  font=\footnotesize,
    legend columns=-1, 
  /tikz/every even column/.append style={column sep=0.2cm}, 
  text opacity=1
},
tick label style={font=\footnotesize},
ylabel style={font=\footnotesize},
tick align=inside,
tick pos=left,
x grid style={gray176},
xmin=-0.63, xmax=7.63,
xtick style={color=black},
xticklabel style={rotate=45, anchor=east},
xtick={0,1,2,3,4,5,6,7},
xticklabels={
  {\{[64,128],16\}},
  {\{[64,128],32\}},
    {\{[64,128],64\}},
  {\{[64,128],128\}},
  {\{[128,256],16\}},
  {\{[128,256],32\}},
  {\{[128,256],64\}},
  {\{[128,256],128\}}
},
y grid style={gray176},
ylabel={Normalized Percentage (\%)},
ymajorgrids,
ymin=20, ymax=100,
ytick style={color=black}
]


\draw[draw=none,fill=blue31119180,postaction={pattern=north east lines,pattern color=black}] (axis cs:-0.3,0) rectangle (axis cs:-0.1,96.6244530285244);
\addlegendimage{ybar,ybar legend,draw=none,fill=blue31119180,postaction={pattern=north east lines,pattern color=black}}
\addlegendentry{Accuracy}

\draw[draw=none,fill=blue31119180,postaction={pattern=north east lines,pattern color=black}] (axis cs:0.7,0) rectangle (axis cs:0.9,96.9601209353509);
\draw[draw=none,fill=blue31119180,postaction={pattern=north east lines,pattern color=black}] (axis cs:1.7,0) rectangle (axis cs:1.9,75.7722838723673);
\draw[draw=none,fill=blue31119180,postaction={pattern=north east lines,pattern color=black}] (axis cs:2.7,0) rectangle (axis cs:2.9,94.4872958001403);
\draw[draw=none,fill=blue31119180,postaction={pattern=north east lines,pattern color=black}] (axis cs:3.7,0) rectangle (axis cs:3.9,96.4372427553466);
\draw[draw=none,fill=blue31119180,postaction={pattern=north east lines,pattern color=black}] (axis cs:4.7,0) rectangle (axis cs:4.9,96.1228197428093);
\draw[draw=none,fill=blue31119180,postaction={pattern=north east lines,pattern color=black}] (axis cs:5.7,0) rectangle (axis cs:5.9,98.5354213812887);
\draw[draw=none,fill=blue31119180,postaction={pattern=north east lines,pattern color=black}] (axis cs:6.7,0) rectangle (axis cs:6.9,97.5444610276115);

\draw[draw=none,fill=orange25512714,postaction={pattern=north west lines,pattern color=black}] (axis cs:-0.1,0) rectangle (axis cs:0.1,95.5637739609773);
\addlegendimage{ybar,ybar legend,draw=none,fill=orange25512714,postaction={pattern=north west lines,pattern color=black}}
\addlegendentry{Specificity}

\draw[draw=none,fill=orange25512714,postaction={pattern=north west lines,pattern color=black}] (axis cs:0.9,0) rectangle (axis cs:1.1,98.3358507606783);
\draw[draw=none,fill=orange25512714,postaction={pattern=north west lines,pattern color=black}] (axis cs:1.9,0) rectangle (axis cs:2.1,95.6087830567541);
\draw[draw=none,fill=orange25512714,postaction={pattern=north west lines,pattern color=black}] (axis cs:2.9,0) rectangle (axis cs:3.1,98.7623600492066);
\draw[draw=none,fill=orange25512714,postaction={pattern=north west lines,pattern color=black}] (axis cs:3.9,0) rectangle (axis cs:4.1,97.2005169809558);
\draw[draw=none,fill=orange25512714,postaction={pattern=north west lines,pattern color=black}] (axis cs:4.9,0) rectangle (axis cs:5.1,94.9191128496039);
\draw[draw=none,fill=orange25512714,postaction={pattern=north west lines,pattern color=black}] (axis cs:5.9,0) rectangle (axis cs:6.1,99.1063392453947);
\draw[draw=none,fill=orange25512714,postaction={pattern=north west lines,pattern color=black}] (axis cs:6.9,0) rectangle (axis cs:7.1,99.1401298681076);

\draw[draw=none,fill=green4416044,postaction={pattern=horizontal lines,pattern color=black}] (axis cs:0.1,0) rectangle (axis cs:0.3,99.4285973051858);
\addlegendimage{ybar,ybar legend,draw=none,fill=green4416044,postaction={pattern=horizontal lines,pattern color=black}}
\addlegendentry{Recall}

\draw[draw=none,fill=green4416044,postaction={pattern=horizontal lines,pattern color=black}] (axis cs:1.1,0) rectangle (axis cs:1.3,93.3230689429462);
\draw[draw=none,fill=green4416044,postaction={pattern=horizontal lines,pattern color=black}] (axis cs:2.1,0) rectangle (axis cs:2.3,23.329943888914);
\draw[draw=none,fill=green4416044,postaction={pattern=horizontal lines,pattern color=black}] (axis cs:3.1,0) rectangle (axis cs:3.3,83.1851995175188);
\draw[draw=none,fill=green4416044,postaction={pattern=horizontal lines,pattern color=black}] (axis cs:4.1,0) rectangle (axis cs:4.3,94.4193552371033);
\draw[draw=none,fill=green4416044,postaction={pattern=horizontal lines,pattern color=black}] (axis cs:5.1,0) rectangle (axis cs:5.3,99.305095281811);
\draw[draw=none,fill=green4416044,postaction={pattern=horizontal lines,pattern color=black}] (axis cs:6.1,0) rectangle (axis cs:6.3,97.0260712771344);
\draw[draw=none,fill=green4416044,postaction={pattern=horizontal lines,pattern color=black}] (axis cs:7.1,0) rectangle (axis cs:7.3,93.3259506568249);
\path [draw=black, semithick]
(axis cs:-0.2,96.5866260101472)
--(axis cs:-0.2,96.6622800469017);

\path [draw=black, semithick]
(axis cs:0.8,96.9241501231867)
--(axis cs:0.8,96.9960917475151);

\path [draw=black, semithick]
(axis cs:1.8,75.6829109134345)
--(axis cs:1.8,75.8616568313);

\path [draw=black, semithick]
(axis cs:2.8,94.4395556928677)
--(axis cs:2.8,94.5350359074129);

\path [draw=black, semithick]
(axis cs:3.8,96.3984247792036)
--(axis cs:3.8,96.4760607314895);

\path [draw=black, semithick]
(axis cs:4.8,96.0824009492529)
--(axis cs:4.8,96.1632385363657);

\path [draw=black, semithick]
(axis cs:5.8,98.5101831638592)
--(axis cs:5.8,98.5606595987182);

\path [draw=black, semithick]
(axis cs:6.8,97.5120117602985)
--(axis cs:6.8,97.5769102949245);

\addplot [semithick, black, mark=-, mark size=2, mark options={solid}, only marks]
table {%
-0.2 96.5866260101472
0.8 96.9241501231867
1.8 75.6829109134345
2.8 94.4395556928677
3.8 96.3984247792036
4.8 96.0824009492529
5.8 98.5101831638592
6.8 97.5120117602985
};
\addplot [semithick, black, mark=-, mark size=2, mark options={solid}, only marks]
table {%
-0.2 96.6622800469017
0.8 96.9960917475151
1.8 75.8616568313
2.8 94.5350359074129
3.8 96.4760607314895
4.8 96.1632385363657
5.8 98.5606595987182
6.8 97.5769102949245
};
\path [draw=black, semithick]
(axis cs:0,95.5131426678483)
--(axis cs:0,95.6144052541063);

\path [draw=black, semithick]
(axis cs:1,98.3042730617697)
--(axis cs:1,98.3674284595869);

\path [draw=black, semithick]
(axis cs:2,95.5583958441253)
--(axis cs:2,95.659170269383);

\path [draw=black, semithick]
(axis cs:3,98.7350267142959)
--(axis cs:3,98.7896933841173);

\path [draw=black, semithick]
(axis cs:4,97.1598887886572)
--(axis cs:4,97.2411451732544);

\path [draw=black, semithick]
(axis cs:5,94.865132722317)
--(axis cs:5,94.9730929768908);

\path [draw=black, semithick]
(axis cs:6,99.0830264496901)
--(axis cs:6,99.1296520410993);

\path [draw=black, semithick]
(axis cs:7,99.1172523325935)
--(axis cs:7,99.1630074036216);

\addplot [semithick, black, mark=-, mark size=2, mark options={solid}, only marks]
table {%
0 95.5131426678483
1 98.3042730617697
2 95.5583958441253
3 98.7350267142959
4 97.1598887886572
5 94.865132722317
6 99.0830264496901
7 99.1172523325935
};
\addplot [semithick, black, mark=-, mark size=2, mark options={solid}, only marks]
table {%
0 95.6144052541063
1 98.3674284595869
2 95.659170269383
3 98.7896933841173
4 97.2411451732544
5 94.9730929768908
6 99.1296520410993
7 99.1630074036216
};
\path [draw=black, semithick]
(axis cs:0.2,99.3978312320774)
--(axis cs:0.2,99.4593633782943);

\path [draw=black, semithick]
(axis cs:1.2,93.2231146679983)
--(axis cs:1.2,93.4230232178941);

\path [draw=black, semithick]
(axis cs:2.2,23.1621787482034)
--(axis cs:2.2,23.4977090296247);

\path [draw=black, semithick]
(axis cs:3.2,83.0359467528689)
--(axis cs:3.2,83.3344522821686);

\path [draw=black, semithick]
(axis cs:4.2,94.3273662735299)
--(axis cs:4.2,94.5113442006767);

\path [draw=black, semithick]
(axis cs:5.2,99.2712717387075)
--(axis cs:5.2,99.3389188249146);

\path [draw=black, semithick]
(axis cs:6.2,96.957772701783)
--(axis cs:6.2,97.0943698524859);

\path [draw=black, semithick]
(axis cs:7.2,93.2260162269535)
--(axis cs:7.2,93.4258850866963);

\addplot [semithick, black, mark=-, mark size=2, mark options={solid}, only marks]
table {%
0.2 99.3978312320774
1.2 93.2231146679983
2.2 23.1621787482034
3.2 83.0359467528689
4.2 94.3273662735299
5.2 99.2712717387075
6.2 96.957772701783
7.2 93.2260162269535
};
\addplot [semithick, black, mark=-, mark size=2, mark options={solid}, only marks]
table {%
0.2 99.4593633782943
1.2 93.4230232178941
2.2 23.4977090296247
3.2 83.3344522821686
4.2 94.5113442006767
5.2 99.3389188249146
6.2 97.0943698524859
7.2 93.4258850866963
};
\end{axis}

\end{tikzpicture} 
        \setlength\abovecaptionskip{0pt}
        \setlength\belowcaptionskip{-3pt} 
        \caption{\centering Performance in a familiar RF environment}
        \label{fig:barplot_perf1}
    \end{subfigure}
    \hfill
        \begin{subfigure}{0.48\textwidth} 
        \centering
        \setlength\fwidth{\linewidth}
        \setlength\fheight{.15\linewidth}
\begin{tikzpicture}

\definecolor{gray176}{RGB}{176,176,176}
\definecolor{gray31119180}{RGB}{31,119,180}
\definecolor{green4416044}{RGB}{44,160,44}
\definecolor{orange25512714}{RGB}{255,127,14}
\definecolor{white204}{RGB}{204,204,204}

\definecolor{gray176}{RGB}{176,176,176}
\definecolor{blue31119180}{RGB}{31,119,180}
\definecolor{green4416044}{RGB}{44,160,44}
\definecolor{orange25512714}{RGB}{255,127,14}
\definecolor{white204}{RGB}{204,204,204}

\begin{axis}[
width=\fwidth,
height=2.5\fheight,
legend cell align={left},
legend style={
  at={(.5,1.35)},
  anchor=north,
  draw=white204,
  fill opacity=0.8,
  draw opacity=1,
  font=\footnotesize,
    legend columns=-1, 
  /tikz/every even column/.append style={column sep=0.2cm}, 
  text opacity=1
},
tick label style={font=\footnotesize},
ylabel style={font=\footnotesize},
tick align=inside,
tick pos=left,
x grid style={gray176},
xmin=-0.63, xmax=7.63,
xtick style={color=black},
xticklabel style={rotate=45, anchor=east},
xtick={0,1,2,3,4,5,6,7},
xticklabels={
  {\{[64,128],16\}},
  {\{[64,128],32\}},
    {\{[64,128],64\}},
  {\{[64,128],128\}},
  {\{[128,256],16\}},
  {\{[128,256],32\}},
  {\{[128,256],64\}},
  {\{[128,256],128\}}
},
y grid style={gray176},
ylabel={Normalized Percentage (\%)},
ymajorgrids,
ymin=20, ymax=100,
ytick style={color=black}
]
\draw[draw=none,fill=blue31119180,postaction={pattern=north east lines,pattern color=black}] (axis cs:-0.3,0) rectangle (axis cs:-0.1,91.1331611478894);
\addlegendimage{ybar,ybar legend,draw=none,fill=blue31119180,postaction={pattern=north east lines,pattern color=black}}
\addlegendentry{Accuracy}

\draw[draw=none,fill=blue31119180,postaction={pattern=north east lines,pattern color=black}] (axis cs:0.7,0) rectangle (axis cs:0.9,94.6514769424988);
\draw[draw=none,fill=blue31119180,postaction={pattern=north east lines,pattern color=black}] (axis cs:1.7,0) rectangle (axis cs:1.9,76.9659786364011);
\draw[draw=none,fill=blue31119180,postaction={pattern=north east lines,pattern color=black}] (axis cs:2.7,0) rectangle (axis cs:2.9,78.9346254356729);
\draw[draw=none,fill=blue31119180,postaction={pattern=north east lines,pattern color=black}] (axis cs:3.7,0) rectangle (axis cs:3.9,92.9575746856904);
\draw[draw=none,fill=blue31119180,postaction={pattern=north east lines,pattern color=black}] (axis cs:4.7,0) rectangle (axis cs:4.9,91.3325872641725);
\draw[draw=none,fill=blue31119180,postaction={pattern=north east lines,pattern color=black}] (axis cs:5.7,0) rectangle (axis cs:5.9,93.42351215615);
\draw[draw=none,fill=gray31119180] (axis cs:6.7,0) rectangle (axis cs:6.9,93.9184182520636);

\draw[draw=none,fill=orange25512714,postaction={pattern=north west lines,pattern color=black}] (axis cs:-0.1,0) rectangle (axis cs:0.1,92.1704307775276);
\addlegendimage{ybar,ybar legend,draw=none,fill=orange25512714,postaction={pattern=north west lines,pattern color=black}}
\addlegendentry{Specificity}

\draw[draw=none,fill=orange25512714,postaction={pattern=north west lines,pattern color=black}] (axis cs:0.9,0) rectangle (axis cs:1.1,98.6226196462537);
\draw[draw=none,fill=orange25512714,postaction={pattern=north west lines,pattern color=black}] (axis cs:1.9,0) rectangle (axis cs:2.1,84.027124918924);
\draw[draw=none,fill=orange25512714,postaction={pattern=north west lines,pattern color=black}] (axis cs:2.9,0) rectangle (axis cs:3.1,80.976482505812);
\draw[draw=none,fill=orange25512714,postaction={pattern=north west lines,pattern color=black}] (axis cs:3.9,0) rectangle (axis cs:4.1,95.0439814308723);
\draw[draw=none,fill=orange25512714,postaction={pattern=north west lines,pattern color=black}] (axis cs:4.9,0) rectangle (axis cs:5.1,91.4033975382059);
\draw[draw=none,fill=orange25512714,postaction={pattern=north west lines,pattern color=black}] (axis cs:5.9,0) rectangle (axis cs:6.1,97.0105744914989);
\draw[draw=none,fill=orange25512714,postaction={pattern=north west lines,pattern color=black}] (axis cs:6.9,0) rectangle (axis cs:7.1,98.3026884423942);
\draw[draw=none,fill=green4416044,postaction={pattern=horizontal lines,pattern color=black}] (axis cs:0.1,0) rectangle (axis cs:0.3,85.8130711869475);
\addlegendimage{ybar,ybar legend,draw=none,fill=green4416044,postaction={pattern=horizontal lines,pattern color=black}}
\addlegendentry{Recall}

\draw[draw=none,fill=green4416044,postaction={pattern=horizontal lines,pattern color=black}] (axis cs:1.1,0) rectangle (axis cs:1.3,74.283738576261);
\draw[draw=none,fill=green4416044,postaction={pattern=horizontal lines,pattern color=black}] (axis cs:2.1,0) rectangle (axis cs:2.3,40.7498084362794);
\draw[draw=none,fill=green4416044,postaction={pattern=horizontal lines,pattern color=black}] (axis cs:3.1,0) rectangle (axis cs:3.3,68.4620703833143);
\draw[draw=none,fill=green4416044,postaction={pattern=horizontal lines,pattern color=black}] (axis cs:4.1,0) rectangle (axis cs:4.3,82.2565271833592);
\draw[draw=none,fill=green4416044,postaction={pattern=horizontal lines,pattern color=black}] (axis cs:5.1,0) rectangle (axis cs:5.3,90.9694058721289);
\draw[draw=none,fill=green4416044,postaction={pattern=horizontal lines,pattern color=black}] (axis cs:6.1,0) rectangle (axis cs:6.3,75.0256975722803);
\draw[draw=none,fill=green4416044,postaction={pattern=horizontal lines,pattern color=black}] (axis cs:7.1,0) rectangle (axis cs:7.3,71.4317752817388);
\path [draw=black, semithick]
(axis cs:-0.2,91.0353876589776)
--(axis cs:-0.2,91.2309346368012);

\path [draw=black, semithick]
(axis cs:0.8,94.5739454644055)
--(axis cs:0.8,94.7290084205921);

\path [draw=black, semithick]
(axis cs:1.8,76.8215564490273)
--(axis cs:1.8,77.1104008237748);

\path [draw=black, semithick]
(axis cs:2.8,78.7947244535381)
--(axis cs:2.8,79.0745264178077);

\path [draw=black, semithick]
(axis cs:3.8,92.86950088183)
--(axis cs:3.8,93.0456484895508);

\path [draw=black, semithick]
(axis cs:4.8,91.2358065398065)
--(axis cs:4.8,91.4293679885385);

\path [draw=black, semithick]
(axis cs:5.8,93.3381674230466)
--(axis cs:5.8,93.5088568892535);

\path [draw=black, semithick]
(axis cs:6.8,93.836106320805)
--(axis cs:6.8,94.0007301833223);

\addplot [semithick, black, mark=-, mark size=2, mark options={solid}, only marks]
table {%
-0.2 91.0353876589776
0.8 94.5739454644055
1.8 76.8215564490273
2.8 78.7947244535381
3.8 92.86950088183
4.8 91.2358065398065
5.8 93.3381674230466
6.8 93.836106320805
};
\addplot [semithick, black, mark=-, mark size=2, mark options={solid}, only marks]
table {%
-0.2 91.2309346368012
0.8 94.7290084205921
1.8 77.1104008237748
2.8 79.0745264178077
3.8 93.0456484895508
4.8 91.4293679885385
5.8 93.5088568892535
6.8 94.0007301833223
};
\path [draw=black, semithick]
(axis cs:0,92.0693330279699)
--(axis cs:0,92.2715285270853);

\path [draw=black, semithick]
(axis cs:1,98.5783282341417)
--(axis cs:1,98.6669110583657);

\path [draw=black, semithick]
(axis cs:2,83.8895825351456)
--(axis cs:2,84.1646673027025);

\path [draw=black, semithick]
(axis cs:3,80.8292060449198)
--(axis cs:3,81.1237589667041);

\path [draw=black, semithick]
(axis cs:4,94.9621486948013)
--(axis cs:4,95.1258141669433);

\path [draw=black, semithick]
(axis cs:5,91.2979417017762)
--(axis cs:5,91.5088533746357);

\path [draw=black, semithick]
(axis cs:6,96.946199848389)
--(axis cs:6,97.0749491346087);

\path [draw=black, semithick]
(axis cs:7,98.2536806224567)
--(axis cs:7,98.3516962623317);

\addplot [semithick, black, mark=-, mark size=2, mark options={solid}, only marks]
table {%
0 92.0693330279699
1 98.5783282341417
2 83.8895825351456
3 80.8292060449198
4 94.9621486948013
5 91.2979417017762
6 96.946199848389
7 98.2536806224567
};
\addplot [semithick, black, mark=-, mark size=2, mark options={solid}, only marks]
table {%
0 92.2715285270853
1 98.6669110583657
2 84.1646673027025
3 81.1237589667041
4 95.1258141669433
5 91.5088533746357
6 97.0749491346087
7 98.3516962623317
};
\path [draw=black, semithick]
(axis cs:0.2,85.5148593061729)
--(axis cs:0.2,86.1112830677221);

\path [draw=black, semithick]
(axis cs:1.2,73.9116704777929)
--(axis cs:1.2,74.6558066747292);

\path [draw=black, semithick]
(axis cs:2.2,40.3341446863203)
--(axis cs:2.2,41.1654721862385);

\path [draw=black, semithick]
(axis cs:3.2,68.0670397162607)
--(axis cs:3.2,68.857101050368);

\path [draw=black, semithick]
(axis cs:4.2,81.9305114547347)
--(axis cs:4.2,82.5825429119837);

\path [draw=black, semithick]
(axis cs:5.2,90.7235997665339)
--(axis cs:5.2,91.2152119777239);

\path [draw=black, semithick]
(axis cs:6.2,74.6571392000012)
--(axis cs:6.2,75.3942559445593);

\path [draw=black, semithick]
(axis cs:7.2,71.0474842089785)
--(axis cs:7.2,71.8160663544992);

\addplot [semithick, black, mark=-, mark size=2, mark options={solid}, only marks]
table {%
0.2 85.5148593061729
1.2 73.9116704777929
2.2 40.3341446863203
3.2 68.0670397162607
4.2 81.9305114547347
5.2 90.7235997665339
6.2 74.6571392000012
7.2 71.0474842089785
};
\addplot [semithick, black, mark=-, mark size=2, mark options={solid}, only marks]
table {%
0.2 86.1112830677221
1.2 74.6558066747292
2.2 41.1654721862385
3.2 68.857101050368
4.2 82.5825429119837
5.2 91.2152119777239
6.2 75.3942559445593
7.2 71.8160663544992
};
\end{axis}

\end{tikzpicture} 
        \setlength\abovecaptionskip{.2pt}
        \setlength\belowcaptionskip{-.2cm} 
        \caption{\centering Performance in an unseen RF environment}
        \label{fig:barplot_perf2}
    \end{subfigure}
\caption{\justifying Model performance (accuracy, specificity, and recall) for Transfer Learning: (a) familiar RF environment and (b) unseen RF environment to evaluate generalization.}
    \label{fig:perf_comparison}
    \Description{Comparison of model performance metrics (accuracy, specificity, and recall) for TL models: (a) when tested in a familiar RF environment and (b) when tested in a previously unseen RF environment to evaluate generalization capabilities}
\end{figure}

Fig.~\ref{fig:perf_comparison} compares accuracy, specificity, and recall for \gls{tl} models, described in Sec.~\ref{subsec:transferlearning}, tested in a familiar \gls{rf} environment (Fig.~\ref{fig:barplot_perf1}), and an unseen \gls{rf} environment (Fig.~\ref{fig:barplot_perf2}), to evaluate generalization capabilities. We select the \{[128,256], 32\} model as the best-performing configuration, based on its strong results in familiar settings (96.12\% accuracy, 94.92\% specificity, and 99.31\% recall) and its superior generalization in the unseen \gls{rf} environment (91.33\% accuracy, 90.97\% specificity, and 91.40\% recall), outperforming the other model configurations.

\vspace{-10pt}
\subsection{Impact of Transfer Learning}
\label{subsec:transferlearning}
We evaluate \dapp performance across various \gls{rf} configurations to assess the impact of the \gls{tl} method on the model.
We use ISEC dataset (Fig.~\ref{fig:map_EXP}b), dominated by High Traffic instances on both sides, as the foundational model for \gls{tl}, then fine-tuned on the EXP dataset (Fig.~\ref{fig:map_EXP}a), detailed in Sec.~\ref{subsec:cnn_model}, which contains sufficient data for all traffic instances, as described in Tab.~\ref{tab:lookup}. Additionally, we train a baseline model using only the EXP dataset, and compare both performances to assess the benefits of \gls{tl}.

Fig.~\ref{fig:conf_matrix_EXP} shows the confusion matrix of the baseline model without \gls{tl}, tested in a familiar \gls{rf} environment using the same pair of \glspl{ru} locations as in the training. In contrast, Fig.~\ref{fig:conf_matrix_Transf1} and Fig.~\ref{fig:conf_matrix_Transf2} present the performance of the \gls{tl} model, evaluated both in the same familiar \gls{rf} setting as the non-\gls{tl} model, and in a new set of \gls{ru} locations, i.e., an unseen environment.
The \gls{tl} model achieves overall better performance in detecting interference when tested in both a familiar (+10.52\%) and unseen (+2.24\%) \gls{rf}. However, it is outperformed by the non-\gls{tl} model in No Traffic scenarios, with performance drops of -2.11\% in the familiar and -5.63\% in the unseen one, likely due to the influence of the foundational model.

\begin{figure}[t]
    \centering
        \begin{subfigure}{0.32\columnwidth}
        \centering
        \setlength\fwidth{1.24\linewidth}
        \setlength\fheight{.8\linewidth}
\begin{tikzpicture}

\definecolor{black38}{RGB}{38,38,38}
\definecolor{gray176}{RGB}{176,176,176}
\definecolor{myred}{RGB}{230,240,255}  
\definecolor{myblue}{RGB}{0,55,120} 

\definecolor{color0}{RGB}{230,240,255}    
\definecolor{color1}{RGB}{198,219,239}    
\definecolor{color2}{RGB}{158,202,225}    
\definecolor{color3}{RGB}{107,174,214}    
\definecolor{color4}{RGB}{66,146,198}     
\definecolor{color5}{RGB}{33,113,181}     
\definecolor{color6}{RGB}{8,81,156}       
\definecolor{color7}{RGB}{0,55,120}       

\begin{axis}[
    width=0.951\fwidth,
    height=1.5\fheight,
    tick align=inside,
    tick pos=left,
    x grid style={gray176},
    xlabel={Predicted label},
    xlabel style={font=\footnotesize},
    xmin=0, xmax=2,
    xtick style={color=black},
    xtick={0.5,1.5},
    xticklabels={INTERF,CLEAN},
    y dir=reverse,
    y grid style={gray176},
    ylabel={True label},
    ylabel style={font=\footnotesize},
    ymin=0, ymax=2,
    ytick style={color=black},
    ytick={0.5,1.5},
    yticklabel style={rotate=90.0},
    yticklabels={INTERF,CLEAN},
    xticklabel style={font=\scriptsize},
    yticklabel style={font=\scriptsize}
]

\fill[color6] (0,0) rectangle (1,1); 
\fill[color1] (1,0) rectangle (2,1); 
\fill[color0] (0,1) rectangle (1,2); 
\fill[color7] (1,1) rectangle (2,2); 


\draw (axis cs:0.5,0.5) node[
    scale=1.1,
    text=white,  
    rotate=0.0,
    align=center
]{\footnotesize 88.73\\
\scriptsize(215527)};

\draw (axis cs:1.5,0.5) node[
    scale=1.1,
    text=black38,  
    rotate=0.0,
    align=center
]{\footnotesize 11.27\\
\scriptsize(27384)};

\draw (axis cs:0.5,1.5) node[
    scale=1.1,
    text=black38,  
    rotate=0.0,
    align=center
]{\footnotesize 2.97\\
\scriptsize(19051)};

\draw (axis cs:1.5,1.5) node[
    scale=1.1,
    text=white,
    rotate=0.0,
    align=center
]{\footnotesize 97.03\\
\scriptsize(623139)};

\end{axis}

\end{tikzpicture} 
        \caption{\centering No TL in familiar RF environment}  
        \label{fig:conf_matrix_EXP}
    \end{subfigure}
    \hfill
    \begin{subfigure}{0.32\columnwidth}
        \centering
        \setlength\fwidth{1.24\linewidth}
         \setlength\fheight{.8\linewidth}
\begin{tikzpicture}
\definecolor{black38}{RGB}{38,38,38}
\definecolor{gray176}{RGB}{176,176,176}
\definecolor{myred}{RGB}{230,240,255}  
\definecolor{myblue}{RGB}{0,55,120} 
\definecolor{color0}{RGB}{230,240,255}    
\definecolor{color1}{RGB}{198,219,239}    
\definecolor{color2}{RGB}{158,202,225}    
\definecolor{color3}{RGB}{107,174,214}    
\definecolor{color4}{RGB}{66,146,198}     
\definecolor{color5}{RGB}{33,113,181}     
\definecolor{color6}{RGB}{8,81,156}       
\definecolor{color7}{RGB}{0,55,120}       
\begin{axis}[
    width=0.951\fwidth,
    height=1.5\fheight,
    tick align=inside,
    tick pos=left,
    x grid style={gray176},
    xlabel={Predicted label},
    xlabel style={font=\footnotesize},
    xmin=0, xmax=2,
    xtick style={color=black},
    xtick={0.5,1.5},
    xticklabels={INTERF,CLEAN},
    y dir=reverse,
    y grid style={gray176},
    ylabel={True label},
    ylabel style={font=\footnotesize},
    ymin=0, ymax=2,
    ytick style={color=black},
    ytick={0.5,1.5},
    yticklabel style={rotate=90.0},
    yticklabels={INTERF,CLEAN},
    xticklabel style={font=\scriptsize},
    yticklabel style={font=\scriptsize}
]
\fill[color7] (0,0) rectangle (1,1); 
\fill[color0] (1,0) rectangle (2,1); 
\fill[color1] (0,1) rectangle (1,2); 
\fill[color7] (1,1) rectangle (2,2); 
\draw (axis cs:0.5,0.5) node[
    scale=1.1,
    text=white,  
    rotate=0.0,
    align=center
]{\footnotesize 99.31\\
\scriptsize (241223)};
\draw (axis cs:1.5,0.5) node[
    scale=1.1,
    text=black38,  
    rotate=0.0,
    align=center
]{\footnotesize 0.69\\
\scriptsize (1688)};
\draw (axis cs:0.5,1.5) node[
    scale=1.1,
    text=black38,  
    rotate=0.0,
    align=center
]{\footnotesize 5.08\\
\scriptsize (32629)};
\draw (axis cs:1.5,1.5) node[
    scale=1.1,
    text=white,
    rotate=0.0,
    align=center
]{\footnotesize 94.92\\
\scriptsize (609562)};
\end{axis}
\end{tikzpicture} 
        \caption{\centering TL in familiar RF environment}  
        \label{fig:conf_matrix_Transf1}
    \end{subfigure}
    \hfill
        \begin{subfigure}{0.32\columnwidth}
        \centering
        \setlength\fwidth{1.24\linewidth}
         \setlength\fheight{.8\linewidth}
\begin{tikzpicture}
\definecolor{black38}{RGB}{38,38,38}
\definecolor{gray176}{RGB}{176,176,176}
\definecolor{myred}{RGB}{230,240,255}  
\definecolor{myblue}{RGB}{0,55,120} 
\definecolor{color0}{RGB}{230,240,255}    
\definecolor{color1}{RGB}{198,219,239}    
\definecolor{color2}{RGB}{158,202,225}    
\definecolor{color3}{RGB}{107,174,214}    
\definecolor{color4}{RGB}{66,146,198}     
\definecolor{color5}{RGB}{33,113,181}     
\definecolor{color6}{RGB}{8,81,156}       
\definecolor{color7}{RGB}{0,55,120}       
\begin{axis}[
    width=0.951\fwidth,
    height=1.5\fheight,
    tick align=inside,
    tick pos=left,
    x grid style={gray176},
    xlabel={Predicted label},
    xlabel style={font=\footnotesize},
    xmin=0, xmax=2,
    xtick style={color=black},
    xtick={0.5,1.5},
    xticklabels={INTERF,CLEAN},
    y dir=reverse,
    y grid style={gray176},
    ylabel={True label},
    ylabel style={font=\footnotesize},
    ymin=0, ymax=2,
    ytick style={color=black},
    ytick={0.5,1.5},
    yticklabel style={rotate=90.0},
    yticklabels={INTERF,CLEAN},
    xticklabel style={font=\scriptsize},
    yticklabel style={font=\scriptsize}
]
\fill[color7] (0,0) rectangle (1,1); 
\fill[color1] (1,0) rectangle (2,1); 
\fill[color1] (0,1) rectangle (1,2); 
\fill[color7] (1,1) rectangle (2,2); 
\draw (axis cs:0.5,0.5) node[
    scale=1.1,
    text=white,  
    rotate=0.0,
    align=center
]{\footnotesize 90.97\\
\scriptsize (48675)};
\draw (axis cs:1.5,0.5) node[
    scale=1.1,
    text=black38,  
    rotate=0.0,
    align=center
]{\footnotesize 9.03\\
\scriptsize (4832)};
\draw (axis cs:0.5,1.5) node[
    scale=1.1,
    text=black38,  
    rotate=0.0,
    align=center
]{\footnotesize 8.60\\
\scriptsize (23592)};
\draw (axis cs:1.5,1.5) node[
    scale=1.1,
    text=white,
    rotate=0.0,
    align=center
]{\footnotesize 91.40\\
\scriptsize (250842)};
\end{axis}
\end{tikzpicture} 
        \caption{\centering TL in unseen RF environment}  
        \label{fig:conf_matrix_Transf2}
    \end{subfigure}
    \setlength\abovecaptionskip{5pt} 
    \caption{\justifying Classification accuracy (percentage, sample count) with and without TL across different RF environments: (a) no TL in a familiar environment, (b) TL tested in a familiar environment, and (c) TL tested in an unseen environment, using the same color bar as Fig.\ref{fig:conf_matrix_nrsim}.}
    \label{fig:conf_matrix_comparison}
\end{figure}

\vspace{-5pt}
\subsection{Timing and Power Benchmarking}
We assess \dapp's robustness by analyzing its inference time, GPU utilization, and power consumption, both with and without it.

\begin{figure}[t]
\vspace{-10pt}
    \centering
\begin{subfigure}{0.49\linewidth}
    \centering
        \setlength\fwidth{\linewidth}
        \setlength\fheight{.5\linewidth}
    \input{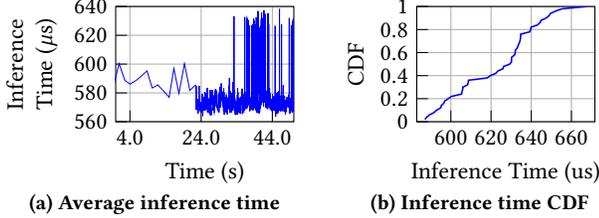} 
    \setlength\abovecaptionskip{0cm} 
    \caption{\centering Average inference time}
    \label{fig:inf_time}
    \Description{Inference Time Variation with Moving Average.}
\end{subfigure}
    \hfill
\begin{subfigure}{0.49\linewidth}
    \centering
        \setlength\fwidth{\linewidth}
        \setlength\fheight{.5\linewidth}
\begin{tikzpicture}

\definecolor{gray176}{RGB}{176,176,176}
\definecolor{white204}{RGB}{204,204,204}

\begin{axis}[
width=0.951\fwidth,
height=1.5\fheight,
legend cell align={left},
legend style={
  fill opacity=0.8,
  draw opacity=1,
  text opacity=1,
  at={(0.03,0.97)},
  anchor=north west,
  draw=white204
},
tick align=inside,
tick pos=left,
x grid style={gray176},
xlabel={Inference Time (us)},
xmajorgrids,
xmin=583.04525, xmax=671.67775,
xtick style={color=black},
y grid style={gray176},
ylabel={CDF},
ymajorgrids,
ymin=0.00, ymax=1.0,
ytick style={color=black}
]
\addplot [semithick, blue]
table {%
587.074 0.02
587.979 0.04
589.421 0.06
591.721 0.08
592.835 0.1
594.919 0.12
595.244 0.14
596.461 0.16
596.997 0.18
598.733 0.2
600.378 0.22
605.362 0.24
605.497 0.26
605.904 0.28
605.978 0.3
608.3 0.32
608.624 0.34
608.761 0.36
618.123 0.38
619.66 0.4
622.375 0.42
623.376 0.44
625.962 0.46
626.247 0.48
629.299 0.5
630.186 0.52
630.186 0.54
630.244 0.56
631.777 0.58
632.001 0.6
632.081 0.62
633.051 0.64
633.465 0.66
633.962 0.68
634.401 0.7
634.603 0.72
634.746 0.74
634.764 0.76
639.182 0.78
639.759 0.8
639.898 0.82
643.119 0.84
644.369 0.86
644.943 0.88
647.437 0.9
648.853 0.92
649.553 0.94
652.063 0.96
655.3 0.98
667.649 1
};
\end{axis}

\end{tikzpicture} 
        \setlength\abovecaptionskip{0cm} 
    \caption{\centering Inference time CDF}
    \label{fig:cdf_time}
    \Description{Inference Time Variation with Moving Average.}
\end{subfigure}
\setlength\abovecaptionskip{-.2cm} 
\setlength\belowcaptionskip{-.2cm} 
    \caption{\justifying Inference time from first iteration: (a) temporal variation with a 5-instance moving average with No (before $24$~s) and High Traffic (after $24$~s), (b) inference time CDF.}
    \label{fig:inference_analysis}
    \Description{GPU utilization (blue) and power draw (orange) comparison with and without \dapp dApp.}
\end{figure}

\textbf{Analysis of Inference Time.}
\label{subsec:inferencetime}
Fig.~\ref{fig:inf_time} shows the temporal variation of the inference time using a 5-instance moving average between No (before $24$~s) and High Traffic (after $24$~s), while Fig.~\ref{fig:cdf_time} presents the corresponding \gls{cdf}.
We observe fluctuations during high-traffic scenarios, when \dapp performs more frequent inferences, likely due to the increased number of operations and contention for \gls{gpu} resources. However, the system maintains overall stability, effectively managing workload distribution.
Additionally, Tab.~\ref{tab:orttimes_model} shows the average inference time for different model configurations, varying with filter counts. We notice a $220$~$\mu$s improvement in the smaller configuration due to the reduced computational complexity.


\begin{table}[t]
    \vspace{10pt}
    \centering
    \captionsetup{aboveskip=.1pt, belowskip=-.27cm} 
    \caption{\justifying Average inference times across different models.}
    \label{tab:orttimes_model}
    \footnotesize
    \begin{tabular}{p{0.4\linewidth} p{0.2\linewidth} p{0.2\linewidth}}  
        \toprule
       \textbf{Model Configuration} & \{[64,128],any\} & \{[128,256],any\} \\
        \midrule
        \textbf{Inference Time ($\mu s$)} & 401.8 & 621.6  \\
        \bottomrule
    \end{tabular}
\end{table}

\begin{figure}[t]
    \centering
    \begin{subfigure}{0.462\textwidth} 
        \centering
                \includegraphics[width=\textwidth]{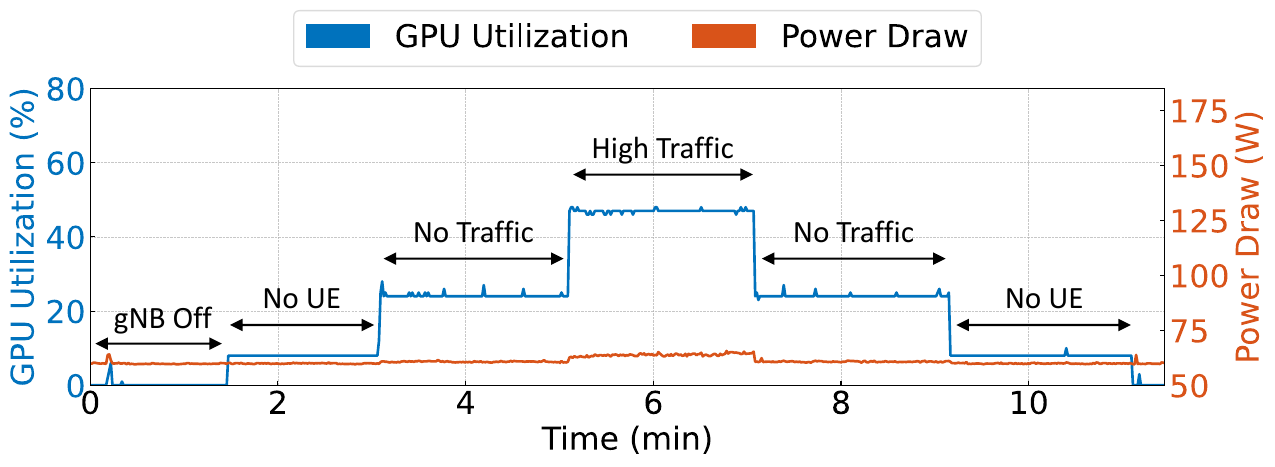}
                \setlength\abovecaptionskip{-.3cm}
  \caption{\centering Without \dapp dApp}
  \label{fig:gpu_nodapp}
    \end{subfigure}
    \hfill
        \begin{subfigure}{0.462\textwidth} 
        \centering
\includegraphics[width=\textwidth]
        {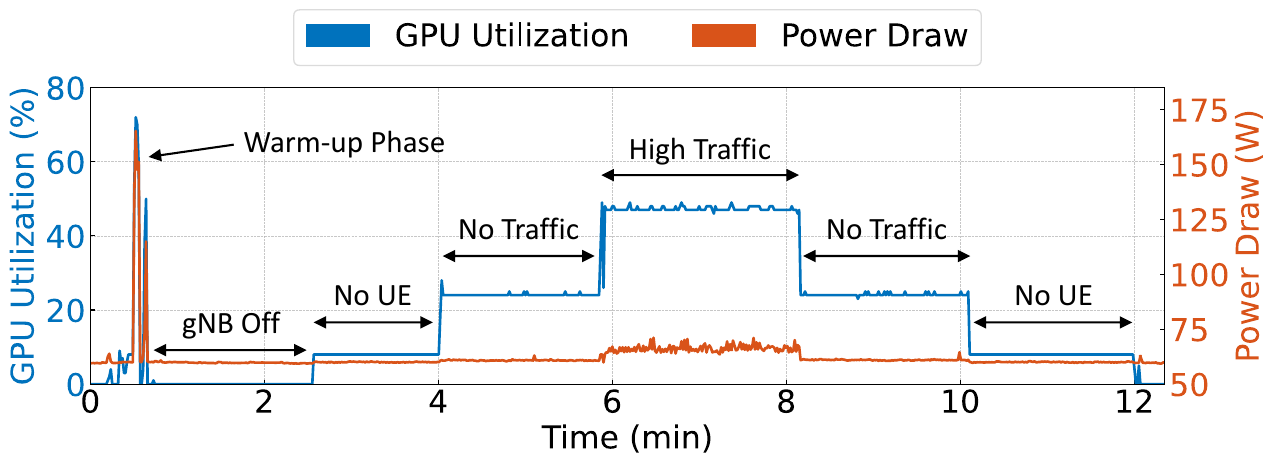}
        \setlength\abovecaptionskip{-.3cm}
  \caption{\centering With \dapp dApp}
  \label{fig:gpu_dapp}
    \end{subfigure}
    \setlength\abovecaptionskip{2pt} 
    \caption{\justifying GPU utilization (blue) and power draw (orange) comparison with and without \dapp dApp.}
    \label{fig:gpu_comparison}
    \Description{GPU utilization (blue) and power draw (orange) comparison with and without \dapp dApp.}
\end{figure}

\textbf{Analysis of GPU Utilization and Power.}
Fig.~\ref{fig:gpu_nodapp} and Fig.~\ref{fig:gpu_dapp} compare A100 \gls{gpu} utilization and power draw with and without \dapp in the same end-to-end scenario. During the warm-up phase (see Sec.~\ref{sec:dappdesign}), \gls{gpu} usage spikes to 21.2\% (vs. 3.3\% without \dapp), and the power rises to $82.73$~W (vs. $61.79$~W). In other phases, power remains similar, except in High Traffic conditions, where it increases from $63.83$~W to $66.13$~W and a higher standard deviation (1.77 vs. 0.67) when using \dapp.
We conclude that \dapp effectively balances the workload without straining \gls{ran} operations, while leveraging \gls{tensorrt} optimizations.

\label{subsec:gpu_usage_power}

\section{Related Work}
\label{sec:relatedwork}

Limited research has been conducted on \gls{ul} interference in \gls{5g} \gls{nr} within real-time timescales and tested on real-world testbed environments.
Existing literature predominantly relies on simulations to predict the presence of interference. For instance, \cite{9860900} and \cite{9475442} detect intermodulation interference in \gls{5g} \gls{nr} using linear regression and \glspl{cnn}, respectively. Similarly, \cite{8885870} and \cite{9600524} detect \gls{lte} \gls{ul} interference using a novel approach that preprocesses time-domain signals into spectral waterfall representations and addresses the problem using an image classification \gls{cnn}. Additionally, authors in \cite{8104767} identify interference among IEEE 802.11b/g, IEEE 802.15.4, and IEEE 802.15.1 using a deep \gls{cnn}.

In the context of spectrum monitoring, \cite{8325299} develops a \gls{cnn} for wireless signal identification, while \cite{9120462} applies logistic regression to detect co-channel interference between \gls{lte} and WiFi users, though both approaches still lack validation beyond simulations. The authors of \cite{uvaydov2021deepsense} propose a framework based on \glspl{cnn} to sense and classify wideband spectrum portions, supported by real-world data collection and testing.
Several studies are conducted to characterize interference signals in various scenarios, such as \cite{s21020597}, which addresses downlink interference in massive \gls{mimo} \gls{5g} macro-cells using geometric channel models, and \cite{testolina2024modeling}, which focuses on modeling interference at higher frequencies for \gls{6g} networks.
Furthermore, \cite{Hattab_2019} and \cite{9860578} discuss interference mitigation mechanisms. The former focuses on angular-based exclusion zones and spatial power control in mmWave frequency ranges, while the latter employs supervised learning-based \gls{iw} selection methods.

Studies on timescales faster than those provided by non-Real-Time (over $1$~sec) and near-Real-Time (between $10$ and $1000$~ms) \glspl{ric} are continuously increasing, highlighting the need for faster and more adaptable control mechanisms capable of responding to dynamic network conditions.
\dapp is based on the O-RAN concept of dApp operating on a timescale faster than $10$~ms, first described in detail in~\cite{D_Oro_2022}. This concept was recently expanded in~\cite{lacava2025dapps}, which presents a comprehensive framework built on the \gls{oai} open-source project and further demonstrates its effectiveness through two use case applications, spectrum sharing and \gls{5g} positioning. Moreover, the authors in~\cite{kouchaki2023openaidapp} extend their previous Open AI Cellular framework to support dApps, validating their approach through experimentation with a federated learning dApp.
Real-time \gls{ran} control is enabled by \cite{bura2024windexrealtimeneuralwhittle} via a Real-Time \gls{ric} and $\mu$App for intelligent MAC-layer scheduling at the \gls{du}. \cite{ko2023edgericempoweringrealtimeintelligent} integrates \gls{ai}-driven decision-making, leveraging \gls{ran} and application data for sub-millisecond latency. CloudRIC \cite{cloudricleo} optimizes \gls{ran} scheduling within O-RAN using heterogeneous computing.
However, to the best of our knowledge, \dapp represents the first end-to-end dApp implementation deployed on a \gls{gpu}-based \gls{ran} node and tested in real-time on a production-ready, real-world testbed.

\vspace{-5pt}
\section{Conclusions and Future Work}
\label{sec:conclusion}

This paper presents \dapp, a \gls{gpu}-accelerated O-RAN dApp that detects in-band \gls{ul} interference with over 91\% accuracy at sub-millisecond speeds (650~$\mu$s). Our solution leverages a \gls{cnn} to analyze \gls{iq} samples directly within the \gls{gnb} physical layer. \dapp is fully implemented in a commercial-grade private 5G network, featuring seamless integration with both NVIDIA Aerial's 5G NR stack and \gls{oai}. The system's effectiveness is validated through extensive evaluations on synthetic and real-world data, confirming its reliability and capabilities.

Future work on \dapp will address several key challenges. We will explore scalability to higher \gls{ue} densities and mobility scenarios, repurpose it for \gls{aoa} estimation/5G positioning, implement mitigation via resource allocation tuning to enhance performance, and develop cross-layer interfaces to enable seamless communication between the dApp and other network components. Finally, we aim to implement online training capabilities allowing the \gls{cnn} to adapt continuously to specific \gls{rf} environments and scenarios.

\vspace{-0.02in}

\section*{Acknowledgements}

This work was partially supported by the U.S. NSF under awards CNS-2112471 and CNS-2434081 and by OUSD(R\&E) through Army Research Laboratory Cooperative Agreement Number W911NF-24-2-0065. The views and conclusions contained in this document are those of the authors and should not be interpreted as representing the official policies, either expressed or implied, of the Army Research Laboratory or the U.S. Government. The U.S. Government is authorized to reproduce and distribute reprints for Government purposes notwithstanding any copyright notation herein.

\vspace{-0.04in}
\balance
\bibliographystyle{IEEEtran}
\bibliography{biblio/sample-base}

\appendix

\end{document}